\documentclass[aps,prl,reprint,twocolumn,superscriptaddress]{revtex4-2}

\usepackage{graphicx}
\usepackage{bm}
\usepackage{hyperref}
\usepackage{amsmath}
\usepackage{amssymb}
\usepackage[table]{xcolor}
\usepackage{array}
\usepackage{multirow,comment}
\usepackage{physics}

\begin{document}

\title{Non-Abelian fractionalization in topological minibands}
\author{Aidan P. Reddy}
\thanks{These authors contributed equally to this work.}
\affiliation{Department of Physics, Massachusetts Institute of Technology, Cambridge, Massachusetts 02139, USA}
\author{Nisarga Paul}
\thanks{These authors contributed equally to this work.}
\affiliation{Department of Physics, Massachusetts Institute of Technology, Cambridge, Massachusetts 02139, USA}
\affiliation{Kavli Institute of Theoretical Physics, University of California, Santa Barbara, Santa Barbara, California 93106, USA}
\author{Ahmed Abouelkomsan}
\affiliation{Department of Physics, Massachusetts Institute of Technology, Cambridge, Massachusetts 02139, USA}
\author{Liang Fu}
\affiliation{Department of Physics, Massachusetts Institute of Technology, Cambridge, Massachusetts 02139, USA}
\date{\today}

\begin{abstract}

Motivated by the recent discovery of fractional quantum anomalous Hall states in moir\'e systems, we consider the possibility of realizing non-Abelian phases in topological minibands. We study a family of moir\'e systems, skyrmion Chern band (SCB) models, which can be realized in two-dimensional semiconductor-magnet heterostructures and also capture the essence of twisted transition metal dichalcogenide (TMD) homobilayers. We show using many-body exact diagonalization that, in spite of strong Berry curvature variations in momentum space, the non-Abelian Moore-Read state can be realized at half filling of the second miniband. These results demonstrate the feasibility of non-Abelian fractionalization in moir\'e systems without Landau levels and shed light on the desirable conditions for their realization. In particular, we highlight the prospect of realizing the Moore-Read state 
in twisted semiconductor bilayers.

\end{abstract}
\maketitle

The fractional quantum Hall (FQH) effect 
has traditionally been limited to the context of two-dimensional electron systems in a strong magnetic field. 
Remarkably, recent experiments have observed a sequence of FQH states in twisted bilayer semiconductor $t$MoTe$_2$ \cite{cai2023signatures,park2023observation, zeng2023thermodynamic,xu2023observation} and rhombohedral pentalayer graphene/hBN \cite{Lu2024Feb} at \textit{zero} field. 
In twisted bilayer semiconductors, the existence of such fractional quantum anomalous Hall (FQAH) states was theoretically predicted \cite{devakul2021magic,li2021spontaneous, crepel2023anomalous} as a consequence of 
Coulomb interactions in partially filled topological moir\'e bands \cite{wu2019topological} and spontaneous time-reversal symmetry breaking. Fractional and integer QAH states have also been proposed in other moir\'e material platforms, including twisted bilayer graphene \cite{Abouelkomsan2020Mar,Ledwith2020May,Repellin2020May, xie2021fractional}, periodically strained graphene \cite{gao2023untwisting, venderbos2016interacting} and narrow gap semiconductors subject to an electrostatic superlattice potential \cite{ghorashi2023topological, tan2024designing}. These theoretical advances and experimental breakthroughs introduce a new frontier of strongly correlated topological quantum matter and offer the potential to achieve 
high-temperature topological protection. 

To date, much work on the FQAH effect has focused on 
filling factors $\nu<1$. Here, theoretical understanding is largely guided by the resemblance between the $|C|=1$ band 
($C$ is the band Chern number) and the lowest Landau level (LLL). 
At fractional filling of a Chern band, Coulomb interaction can drive the system into fractional Chern insulator (FCI) states--the lattice analogs of FQH states \cite{sheng2011fractional,neupert2011fractional,regnault2011fractional,tang2011high,sun2011nearly}. A natural question is how far this guiding principle can be pushed to higher fillings.

While the LLL typically hosts Abelian topological orders at fractional fillings, the first excited Landau level (1LL) is predicted to host even richer \textit{non}-Abelian topological orders \cite{Willett1987Oct}, e.g. the phase of the Moore-Read Pfaffian/ anti-Pfaffian state \cite{Moore1991Aug,Read2000Apr,Levin2007Dec} or the Read-Rezayi state \cite{Read1999Mar}. These phases support fractional quasiparticles obeying non-Abelian exchange statistics \cite{Witten1989Sep,Fredenhagen1989Jun,Frohlich1990Jan,Imbo1990Jan} and could provide a platform for fault-tolerant quantum computation \cite{Kitaev2003Jan,Freedman2003,Nayak2008Sep}. 
Theoretical studies have also explored non-Abelian phases of fermions and bosons in lattice models \cite{bernevig2012emergent,liu_non-Abelian_2013, wang2015fermionic} as well as quantum spin systems \cite{Kitaev2006Jan, Jackeli2009Jan}. 
The advent of FQAH materials raises the exciting prospect of realizing non-Abelian fractionalization without a magnetic field and at elevated temperatures. \par

As a starting point, we note that theory predicts that the second moir\'e band in twisted transition metal dichalcogenide ($t$TMD) homobilayers is flat, well isolated, and has the same sign Chern number as the first moiré band over a wide range of twist angles \cite{devakul2021magic, reddy2023fractional}. 
Motivated by this resemblance to the 1LL, we consider the possibility of non-Abelian FQAH states in the second $t$TMD miniband. 
In particular, we study a family of continuum models, \textit{skyrmion Chern band} models, which capture in a minimal setting the essential features of topological minibands in twisted TMD bilayers (such as $t$MoTe$_2$ and $t$WSe$_2$) as well as other material platforms. Within this family of models, including an ``adiabatic model" for $t$MoTe$_2$ \cite{zhai2020theory,morales2023magic},
we establish using many-body exact diagonalization that the analog of the non-Abelian state 
in the half-filled $n=1$ LL can be realized in the second topological miniband. Our study provides a realistic material proposal for realizing non-Abelian phases in topological minibands and sheds light on the desirable conditions for their realization. \par 

\begin{figure*}
\centering
\includegraphics[width=\linewidth]{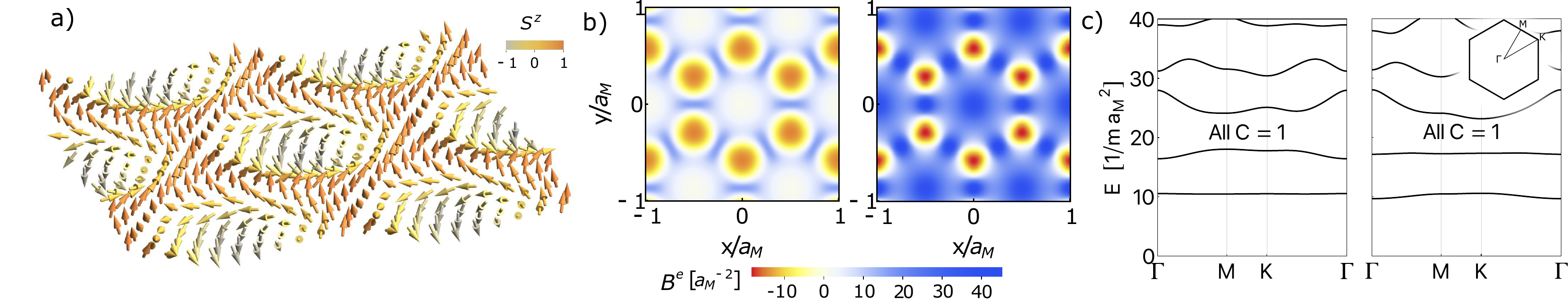}
\caption{\textbf{Skyrmion Chern band (SCB) models.} Electrons strongly coupled to a skyrmion-like spin or pseudospin texture may exhibit flat topological minibands. (a) An example texture ($\alpha=1, N_0=0.11$). (b) Emergent magnetic field $B^e$, or scalar chirality, associated with 
skyrmion textures (from part (a), on left; with $\alpha=1.5, N_0=0.4$, on right). (c) Respective electronic band structures (using Eq. \eqref{eq:H}), which exhibit flat 1st or 2nd topological minibands. Energy measured from $-J$ at large $J$.
}
\label{fig:skx}
\end{figure*}

The following simple Hamiltonian defines our family of skyrmion Chern band (SCB) models in two dimensions:

\begin{equation}\label{eq:H}
  H = \frac{p^2}{2m} +J \bm \sigma \cdot \bm S({\bm r}).%
\end{equation}
Here, $\bm \sigma$ is the spin (or pseudospin) $\frac12$ operator that couples to a periodic Zeeman field $\bm S({\bm r})= \bm S({\bm r + \bm a_{1,2}})$ with strength $J$.
We assume $\bm S(\bm r)$ is a noncoplanar (chiral) texture that defines 
a mapping of the real-space unit cell 
(a torus) to the Bloch sphere with a nonzero winding number, such as a \textit{skyrmion} texture. 
It is well known that chiral textures give rise to an emergent real-space magnetic field $B^e(\bm r) = -\frac{\hbar}{2e} \hat{\bm S} \cdot (\partial_x \hat{\bm S} \times \partial_y \hat{\bm S})$, with an integer number of flux quanta per unit cell ($\hat{\bm S} = \bm S/S$). 
The resulting ``topological Hall effect" has been widely studied in magnetic metals \cite{bruno2004topological,Han2017}. On the other hand, skrymion textures have been less studied in low-density semiconductors. Here, the topological Hall effect manifests as a quantized anomalous Hall effect due to the formation of Chern bands \cite{Hamamoto2015Sep}. 
In a recent work \cite{paul2023giant}, we showed that remarkably, these Chern bands can become \textit{flat} at a magic value of the magnetization $\overline{m}\equiv \int_{\text{u.c.}} S^z(r)/A_{uc}$, where $A_{uc}$ is the unit cell area. SCB models can be potentially realized in a two-dimensional (2D) semiconductor proximity-coupled to a magnetic insulator. 
Ref. \cite{paul2023giant} proposes the heterostructure MoS2/CrBr3, which has the following advantages: large exchange energy J, and the likelihood of a skyrmion crystal (SkX) even at zero field in chromium trihalide bilayers \cite{Akram2021Aug,Hejazi2020May,Hejazi2021Sep}. \par

As we now describe, the combination of large $J$ ($J \gg \hbar^2/ma^2$, where $a$ is the moir\'e period and $m$ is the effective mass of charge carriers) and an SkX allows for flat Chern bands. The large-$J$ or ``adiabatic" limit, which is generally achievable \cite{paul2023giant}, 
enforces local alignment of the electron spin to $\bm S({\bm r})$, which in turn induces a Berry phase. This is made clear by a position-dependent 
unitary transformation $\bm U({\bm r})$ which rotates the spin texture $\bm{S}(\bm{r})$ into $S({\bm r}) \hat z$ and introduces a gauge field $\bm A_i = \frac{i\hbar}{e} \bm U^\dagger \partial_i \bm U.$ For large $J$, we make an ``adiabatic" approximation and project 
onto the low-energy manifold of locally spin (anti)-aligned electrons and obtain the effective adiabatic Hamiltonian \cite{Bruno2004Aug,paul2023giant}
\begin{equation}\label{eq:Heff}
  H_{\text{ad}} = \frac{(\bm p- e\bm A(\bm r))^2}{2m} +\frac{\hbar^2}{8m} (\partial_i \hat{\bm S})^2 - J S({\bm r})
\end{equation}
(summing $i=x,y$), where now $\bm A(\bm r)$ is the $\downarrow\downarrow$ component of the SU(2) gauge field and the second term originates from its off-diagonal elements. We refer to $\bm A$ as the emergent gauge field, with curl $B^e(\bm r) =-\frac{\hbar}{2e}\hat{ \bm S} \cdot (\partial_x \hat{\bm S} \times \partial_y\hat{ \bm S})$, which is generally nonuniform. A skyrmion texture is one with a single flux quantum of emergent magnetic field $B^e$ per unit cell, and is the simplest topologically nontrivial periodic spin texture. Eq. \eqref{eq:Heff} establishes a parallel with ordinary Landau levels. The full SCB Hamiltonian Eq. ~\ref{eq:H} describes (pseudo-)spinful electrons without a magnetic field while the adiabatic Hamiltonian Eq. ~\ref{eq:Heff} describes (pseudo-)spinless electrons in a magnetic field. However, all low-energy physical quantities associated with the two Hamiltonians are identical in the limit $J/(\hbar^2/(ma^2)) \rightarrow \infty$.
\par

Indeed, TMD homobilayers are approximate SCB models with the role of $\bm \sigma$ played by the layer degree of freedom and their Chern bands can be understood in a similar fashion \cite{wu2019topological,zhai2020theory,morales2023magic}. 
In this case, the layer-pseudospin skyrmion texture
corresponds to interlayer tunnelings and intralayer potentials within the $K$ and $K'$ valleys that vary spatially according to local interlayer stacking \cite{wu2019topological}. Over a wide range of fillings, charge carriers are driven into one valley by Coulomb 
interactions, spontaneously breaking time-reversal symmetry \cite{crepel2023anomalous,Anderson2023Jul, reddy2023fractional}. 
The lowest band can be made flat by tuning twist angle \cite{devakul2021magic,morales2023magic}, which creates favorable 
conditions for FQAH states \cite{li2021spontaneous,crepel2023anomalous, reddy2023fractional,morales2023pressure,Wang2023Apr,reddy2023toward}. \par

We now proceed to study the model defined by Eq. \eqref{eq:Heff}. 
For the skyrmion texture, we adopt a simple ansatz built out of three harmonics. In particular, we take $\bm S({\bm r})= \bm N({\bm r})/ N({\bm r})$ with 
\begin{equation}\label{eq:skx}
  \bm N(\bm r) =\frac{1}{\sqrt{2}}\sum_{j=1}^6 e^{i \bm q_j\cdot \bm r}\hat{\bm e}_j + N_0 \hat z
\end{equation}
where $\bm q_j = \frac{4\pi}{\sqrt{3}a}(\cos \theta_j,\sin\theta_j)$ and $\hat {\bm e}_j=(i\alpha\sin\theta_j,-i\alpha\cos\theta_j,-1)/\sqrt{2}$ and the angles satisfy $\theta_2=\theta_1+2\pi/3,\theta_3=\theta_1+4\pi/3$, and $\theta_{j+3}=\theta_j+\pi$. This texture can be thought of as a normalized sum of three spin spirals forming a triangular SkX, as plotted in Fig. \ref{fig:skx}a. It is widely adopted in studies of chiral magnets and magnetic skyrmion crystals and qualitatively reproduces the real-space images of skyrmion crystals \cite{Karube2017Dec,Park2011May,Tokura2020Nov,Lin2016Feb,Tonomura2012Mar}. The parameter $\alpha$ controls coplanarity, while $N_0$ controls the out-of-plane magnetization $\overline{m}$ (which monotonically increases with $N_0$). 
We will refer to the Hamiltonian defined with the skrymion texture of Eq. \ref{eq:skx} as the SkX model.

\begin{figure}
  \centering
\includegraphics[width=\columnwidth]{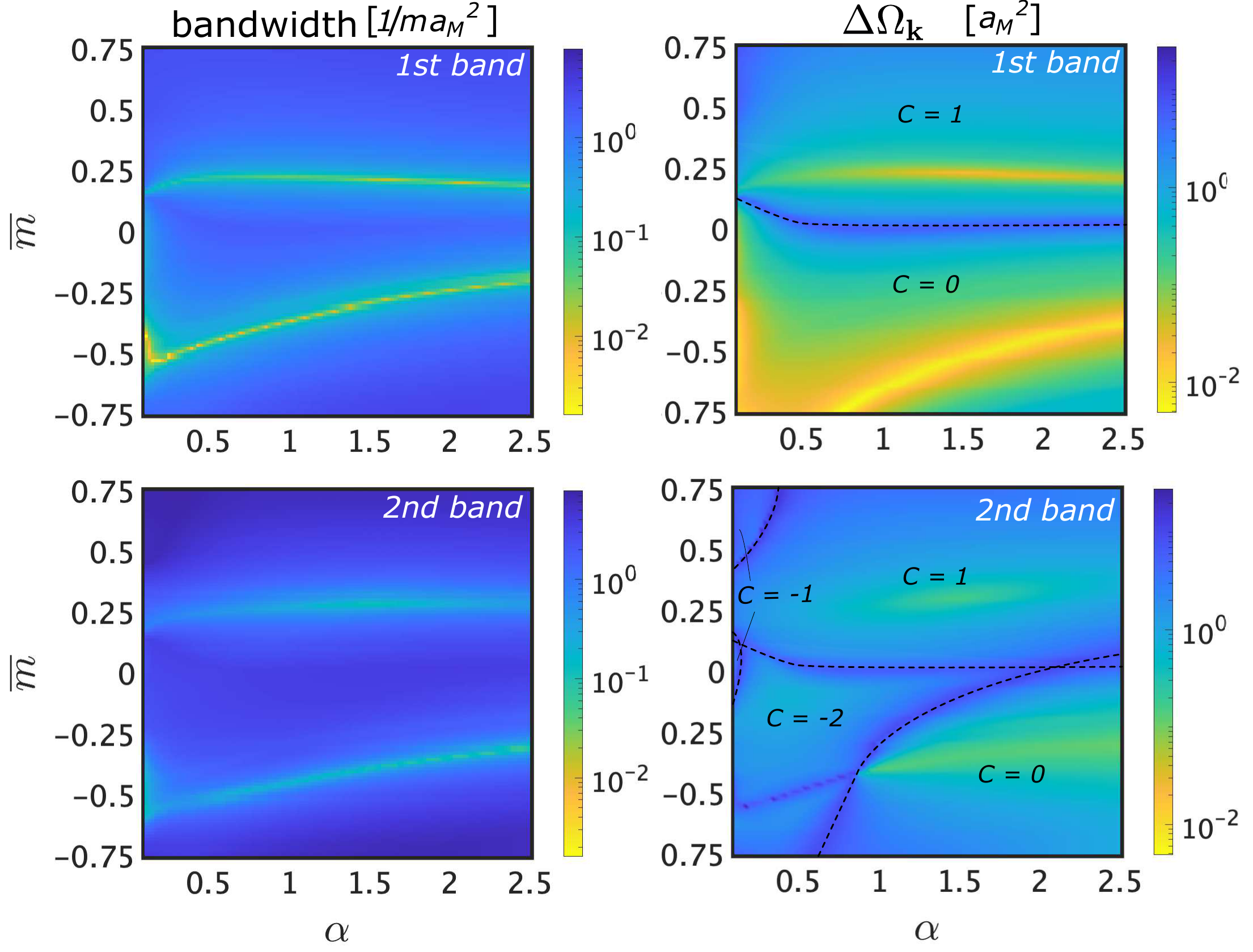}
  \caption{\textbf{Band properties.} (Left) Bandwidths and (right) Berry curvature standard deviations of the two lowest bands in the SkX 
  model as a function of coplanarity $\alpha$ and average $S^z$-magnetization $\overline{m}$, calculated using Eq. \eqref{eq:H}. Bands exhibit minima in similar regions.}
\label{fig:bandproperties}
\end{figure}

In Fig. \ref{fig:skx}b we plot the emergent magnetic field for two examples textures, along with their associated large-$J$ bandstructures in Fig. \ref{fig:skx}c. 
Even when $B^e(\bm r)$ is highly nonuniform and changes sign within the unit cell, we observe that the topological minibands can be made quite flat. This is corroborated in Fig. \ref{fig:bandproperties}, where we plot bandwidths and Berry curvature standard deviations for the lowest two bands across a range of parameters $\alpha$ and $\overline{m}$. Both quantities have a ``magic line" in this parameter space, meaning they can be made relatively small by tuning a single parameter. Remarkably, we will see that
a non-Abelian state can occur
even when the Berry curvature fluctuates wildly.
 \par 

We now study many-body physics at filling $\nu=\frac{3}{2}$ (that is, half-filling of the second miniband assuming full spin polarization) via numerical diagonalization. From here on, to enable direct comparison with Landau levels, we work with the effective Hamiltonian, Eq. ~\ref{eq:Heff}. 
To make the many-body calculation tractable, we restrict our variational Hilbert space to that in which $N_{uc}$ electrons fill the lowest band and $N_e -N_{uc}$ electrons remain in the second miniband where $N_{e}$ is the number of electrons. 
The full lowest band produces a Hartree-Fock self-energy $\Sigma(\bm{k})$ for particles in the second band that is accounted for with a renormalized single-particle energy dispersion, $\tilde{\varepsilon}(\bm{k}) = \varepsilon(\bm{k})+\Sigma(\bm{k})$ \cite{supp}. Here, $\varepsilon(\bm{k})$ is the non-interacting dispersion of the second miniband. Explicitly, we numerically diagonalize the effective projected Hamiltonian
\begin{align}\label{eq:manyBodyProjHam}
  \begin{split}
    \bar{H} &= \sum_{\bm{k}\in BZ}\tilde\varepsilon(\bm{k})n(\bm{k}) + \frac{1}{2A}\sum_{\bm{q} \neq 0}v(\bm{q}):\bar{\rho}(-\bm{q})\bar{\rho}(\bm{q}):
  \end{split}
\end{align}
where $n(\bm{k})=c_{2,\bm{k}}^{\dag}c_{2,\bm{k}}$, $\bar{\rho}(\bm{q}) = P_2\sum_{i}e^{-i\bm{q}\cdot\bm{r}_i}P_2$, ``$::$" orders $c^{\dag}_{2,\bm k}$'s to the left of $c_{2,\bm k}$'s, and $v(\bm{q})=\frac{2\pi e^2}{\epsilon|\bm{q}|}$ is the Fourier transform of the Coulomb potential. $P_2$ is a projector onto the Fock space of the second miniband and $c^{\dag}_{2,\bm{k}}$ creates a (magnetic) Bloch state in the second miniband. In the Supplemental Material, we describe our method for diagonalizing Eq. \ref{eq:Heff} in a Landau Level basis and define all finite-size clusters we use, along with other methodology \cite{supp}.



\begin{figure}
\centering
\includegraphics[width=0.8\columnwidth]{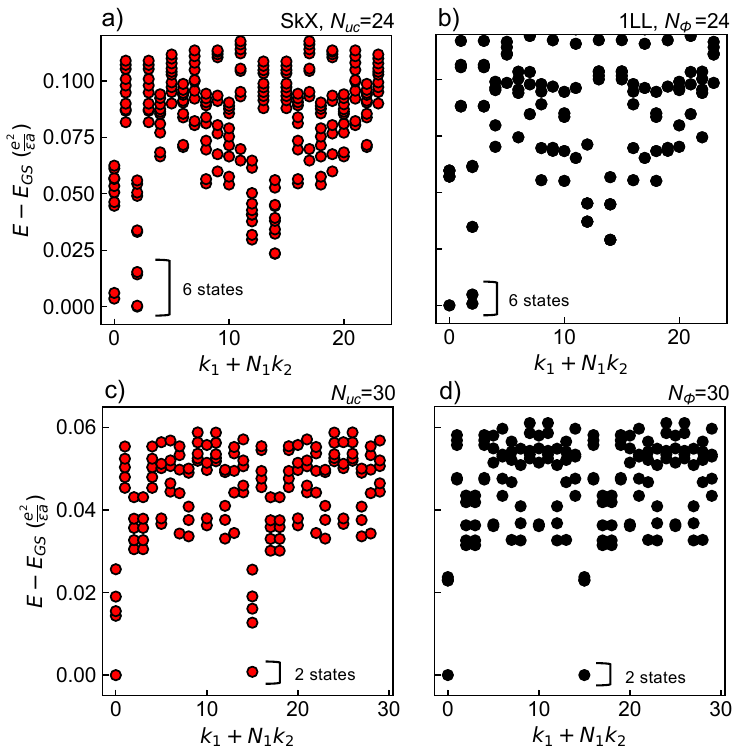}
\caption{\textbf{Non-Abelian states in SkX model. }
(a) Low-energy many-body spectrum of the effective projected Hamiltonian Eq. ~\ref{eq:manyBodyProjHam} at half-filling of the second miniband computed via numerical diagonalization on cluster 24A, and (b) corresponding calculation of the 1LL. (c,d) Analogous data from cluster 30. 
The 10 (a,b) and 5 (c,d) lowest energy levels in each momentum sector are shown. $\alpha=1$, $N_0=0.28$ (equivalently, $\overline{m}=0.271$) and we set $\sqrt{\frac{4\pi}{\sqrt{3}}}\frac{e^2}{\epsilon a} = \frac{2\pi\hbar^2}{mA_{uc}}$.
}
\label{fig:threeHalvesED}
\end{figure}


In Fig. ~\ref{fig:threeHalvesED}, we show many-body spectra obtained by diagonalizing $\bar{H}$ on two finite-size toruses. For $N_{uc}=24$, the number of electrons occupying the second miniband is even (12), while for $N_{uc}=30$, it is odd (15). In the two cases, we observe sixfold and two-fold ground state quasi-degeneracies. (The finite-size splitting among these quasi-degenerate states decreases with increasing interaction strength.) 
These are precisely the degeneracies expected of a Moore-Read state on the torus due to an even-odd effect \cite{Read2000Apr,Oshikawa2007Jun,papic2012quantum}. 
Moreover, the quasi-degenerate ground states have the same center-of-mass momenta as both the Coulomb ground state of the 1LL and the Moore-Read state \cite{supp,peterson2008finite, wang2009particle}. We note that the Hartree-Fock self-energy $\Sigma(\bm{k})$ seems to help stabilize these ground state quasi-degeneracies -- upon neglecting $\Sigma(\bm{k})$, they may not appear for a given parameter set.

Having established a Moore-Read state in the SkX model, we now turn to twisted TMDs. The layer pseudospin texture in $t$TMDs differs from the spin texture of the SkX model: it has different spatial symmetry and contains two merons (half skyrmions) rather than a single skyrmion per unit cell. Additionally, its emergent magnetic field comprises sharp peaks forming a kagome lattice and its pseudospin field strength $S(\bm r)$ is non-uniform. 

\begin{figure}
\centering
\includegraphics[width=\columnwidth]{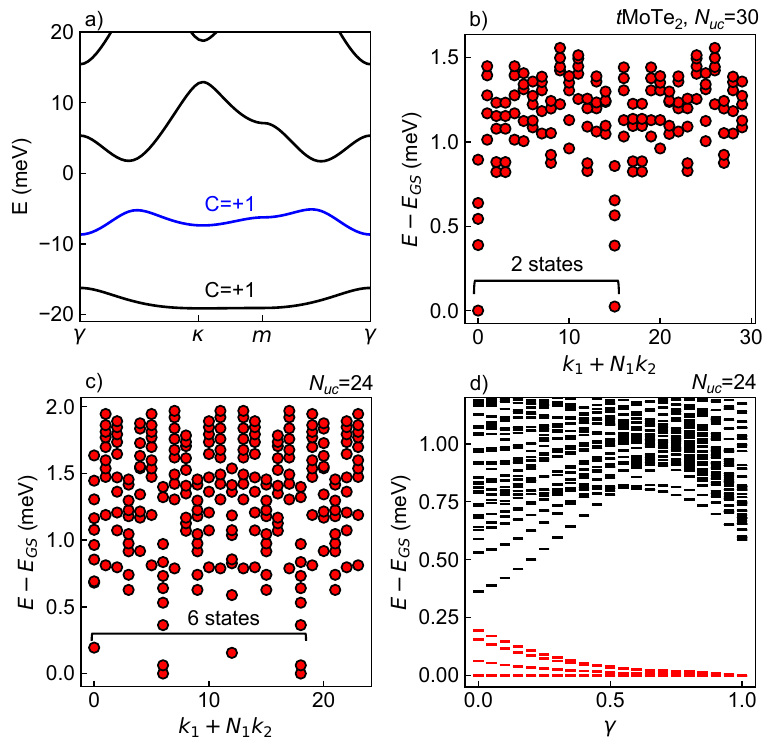}
\caption{\textbf{Adiabatic model for $t$MoTe$_2$.} (a) Minibands of holes in $t$MoTe$_2$ calculated in the adiabatic model at $\theta=2.5^{\circ}$ with parameters from Ref. \cite{reddy2023fractional}. 
Exact diagonalization spectra at $\nu=\frac{3}{2}$ assuming full spin polarization on clusters 30 (b) and 24B (c), using Coulomb interaction with $\epsilon=5$, showing the expected topological ground state quasi-degeneracies of the Moore-Read state. (d) Spectrum of Pf interpolation Hamiltonian $H_{\gamma}$ on cluster 24B with six quasidegenerate ground states highlighted in red.}
\label{fig:TMDFig}
\end{figure}

Despite these differences, our calculations for $t$MoTe$_2$ (using the adiabatic model for the $t$MoTe$_2$, which we detail in the Supplemental Material \cite{supp}) find the same ground state quasidegeneracies at half filling of its second miniband 
(Fig. ~\ref{fig:TMDFig}(b-c)). 
In Fig. \ref{fig:TMDFig}(d), we show that the ground state evolves continuously to the Moore-Read model wavefunction under an interpolation Hamiltonian \cite{supp}, $H_{\gamma} = (1-\gamma)H_{\text{TMD}} + \gamma H_{\text{Pf}}$, preserving the gap throughout the range $0\leq \gamma \leq 1$. Here, $H_{\text{Pf}}$ is the Pfaffian parent Hamiltonian \cite{rezayi2000incompressible,supp}. This strongly supports the non-Abelian nature of the ground state in $t$MoTe$_2$.


\par 

In the half-filled first Landau level, because of exact particle-hole (PH) symmetry, the Pfaffian (Pf) state and its PH conjugate (aPf) are degenerate in the thermodynamic limit \cite{lee2007particle,levin2007particle}. In contrast, our system lacks PH symmetry, so this degeneracy is not expected. To distinguish the two candidate topological orders, we compare overlaps of the $t$MoTe$_2$ ground state with the exact Pf and aPf model states. For the same parameters as in Fig. \ref{fig:TMDFig}, 
we find that the $t$MoTe$_2$ ground state on cluster 26 (24A, 24B) has overlaps of .67 $(.63, .64)$ with the Pf ground state subspace and .58 $(.55,.49)$ with the aPf subspace \cite{supp}.
The enhancement of Pf relative to aPf overlaps suggests that Pf topological order is more likely to emerge in the thermodynamic limit. We note that band mixing effects \cite{abouelkomsan2024band} may alter the ground state.

\begin{figure}
\centering
\includegraphics[width=\columnwidth]{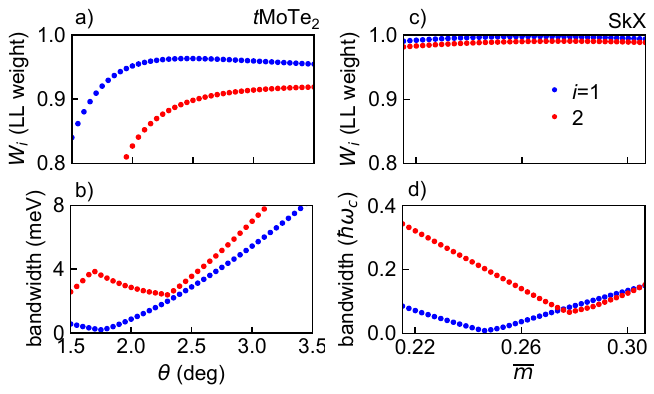}
\caption{\textbf{LL weight and bandwidth.} (a) LL weight $W_i$ (Eq. ~\ref{eq:LLWeight}) and (b) bandwidth of the lowest two minibands of the adiabatic model for $t$MoTe$_2$ with parameters from Ref. \cite{reddy2023fractional}. (c,d) Analogous quantities for the SkX model at fixed $\alpha=1$ as a function of average spin texture magnetization $\overline{m}$. In (d), $\hbar\omega_c = 2\pi\hbar^2/(m A_{uc})$ is the effective cyclotron energy.}
\label{fig:LLWeight}
\end{figure}

We now address the question of why SCB models host non-Abelian FQAH states resembling that of the half-filled 1LL. 
A natural way to quantify the similarity between the $i^{th}$ SCB miniband and the corresponding LL is through 
the ``LL weight",
\begin{align}\label{eq:LLWeight}
  \begin{split}
    W_i = \frac{1}{N_{uc}}\sum_{\bm{k}}\left|\bra{\psi_{\text{ad}}
    ^{(i)}(\bm{k})}\ket{\psi_{\text{LL}}^{(i-1)}(\bm{k})}\right|^2.
  \end{split}
\end{align}
Here, $\ket{\psi_{\text{ad}}^{(i)}}$ and $\ket{\psi^{(n)}_{\text{LL}}}$ are magnetic Bloch states of the $i^{th}$ SCB band within the adiabatic approximation and the $n$LL respectively.Despite the magnetic field and scalar potential fluctuations present in the effective Hamiltonian Eq. \ref{eq:Heff}, 
its LL weights turn out quite high. In Fig. \ref{fig:LLWeight}(a,b), we show the LL weights of the lowest two minibands in the adiabatic model for $t$MoTe$_2$. 
Over a broad range of twist angles the LL weights of the first and second minibands remain high, $> 90 \%$. While the LL weights vary smoothly with interlayer twist $\theta$, the bandwidths show cusp minima at ``magic" values of $\theta$ \cite{devakul2021magic,paul2023giant,morales2023magic}. At small twist angles, the low-energy wavefunctions tend to localize in real space about maxima in $S(\bm{r})$ and form bands with low LL weights. 
At large twist angles, LL weights become larger but the adiabatic approximation becomes less accurate because kinetic energy increases relative to $J$. The lowest two bands of the SkX model exhibit even higher LL weights than the $t$TMD adiabatic model, 
$>99\%$ (Fig. ~\ref{fig:LLWeight}(c,d)).


Remarkably, we find that even when the LL weight is very close to 1, Berry curvature can vary strongly throughout the Brillouin zone. For instance, given the SkX model parameters of ~Fig. \ref{fig:threeHalvesED}, the LL weight of the second band is $0.989$, yet the Berry curvature, in units such that its average is unity, ranges from 0.16 to 1.53 
with a standard deviation of 0.38. The Berry curvature in the adiabatic $t$MoTe$_2$ model at $\theta=2.5^{\circ}$ fluctuates even more strongly, varying from $-3.4$ to $+4.3$ with a standard deviation of 1.58. 
Notably, the quantum weight $K=\frac{1}{2\pi} \int\, d^2 k \Tr(g(\bm{k}))$, where $g(\bm{k})$ the Fubini-Study metric \cite{onishi2024quantum, onishi2023quantum}, is $\sim 3.02$ for the SkX model and $\sim 3.23$ for the adiabatic $t$MoTe$_2$ model-- both close to the value of the 1LL (3 and 2$n$+1 for the $n$LL \cite{ozawa2021relations}).

\par

In conclusion, we have shown that a family of 
skyrmion Chern band models 
with two-body Coulomb interactions host Moore-Read fractional quantum anomalous Hall states. SCB models are naturally realized in 2D semiconductors proximity-coupled to 2D magnets \cite{paul2023giant}, and approximate the TMD moiré systems in which Abelian FQAH states have been observed \cite{wu2019topological,cai2023signatures,zeng2023thermodynamic, park2023observation,xu2023observation}. 
At filling $\nu=\frac{3}{2}$, our many-body exact diagonalization spectra exhibit the topological degeneracy expected of the Moore-Read state, including the even-odd effect \cite{Oshikawa2007Jun}.


 
\par 

The realization of non-Abelian topological order in any setting carries with it significant challenges and equally significant reward. In moir\'e systems, these challenges could be mitigated by the exceptional degree of tunability. Moreover, our work shows that the necessary conditions are far less stringent than might have been thought, and band properties can deviate significantly from those of the 1LL. For instance, Berry curvature can be \textit{far} from uniform. 

Finally, we note that there is evidence for nontrivial topology and ferromagnetism in the second miniband of $t$TMDs. In particular, \emph{double} quantum spin Hall states at $\nu=4$ in $t$MoTe$_2$ and $t$WSe$_2$ \cite{devakul2021magic,kang2024observationMoTe2,kang2024observationWSe2}, enabled by time-reversal symmetry and robust spin conservation, indicate that the first two moiré bands of a given spin have Chern numbers of the same sign \cite{wu2019topological,reddy2023fractional}. 
Further, small-angle $t$MoTe$_2$ devices show large anomalous Hall resistance in the range $1 < \nu < 2$ \cite{kang2024observationMoTe2}, indicating spontaneous spin polarization. Full spin polarization may enable a non-Abelian FCI state at filling factor $\nu=\frac{3}{2}$, as in monolayer WSe$_2$ under a large magnetic field \cite{shi2020odd,pack2023charge}. In addition, a non-Abelian state may occur at $\nu=\frac{5}{2}$, due to complete filling of the lowest miniband of both spins and half filling of the second miniband of one spin.


\textit{Note added} — Several other papers discussing non-Abelian states in $t$MoTe$_2$ have appeared recently \cite{xu2024multiple,ahn2024first,wang2024higher,chen2024robust}.

\begin{acknowledgments}
\emph{Acknowledgements--} We thank Yang Zhang for collaboration on Ref.\cite{paul2023giant}. We thank Kin Fai Mak, Jie Shan, Kaifei Kang, Emil Bergholtz and Xiaodong Xu for stimulating discussions. 
A.P.R. acknowledges helpful conversations with Nicolás Morales-Durán, Sankar Das Sarma, and Kirill Shtengel. N.P. acknowledges helpful conversations with Tixuan Tan. This work was supported by the Air Force Office of Scientific Research (AFOSR) under Award No. FA9550-22-1-0432. 
The authors acknowledge the MIT SuperCloud and Lincoln Laboratory Supercomputing Center for providing computing resources that have contributed to the research results reported within this paper. N.P. was supported in part by grant NSF PHY-2309135 to the Kavli Institute for Theoretical Physics (KITP).
A.A was supported by the Knut and Alice Wallenberg Foundation (KAW 2022.0348). L.F. was partly supported by the Simons Investigator Award from the Simons Foundation.  
\end{acknowledgments}

\end{document}


\title{Supplemental Material for \emph{Non-Abelian fractionalization in topological minibands}}
\author{Aidan P. Reddy}
\thanks{These authors contributed equally to this work.}
\author{Nisarga Paul}
\thanks{These authors contributed equally to this work.}
\author{Ahmed Abouelkomsan}
\author{Liang Fu} 
\affiliation{Department of Physics, Massachusetts Institute of Technology, Cambridge, MA,
USA}

\maketitle
\tableofcontents

\section{Pfaffian wavefunction and parent three-body interaction}
\subsection{Proof that Pfaffian state is a zero-energy eigenstate of parent interaction}
The Pfaffian wavefunction for an even number of electrons on a plane is \cite{Moore1991Aug}
\begin{align}
    \begin{split}
        \Psi_{\text{Pf}}(\{z_i\}) = \text{Pf}\left(\frac{1}{z_i-z_j}\right)\prod_{i<j}(z_i-z_j)^2e^{-\frac{1}{4\ell^2}\sum_i|z_i|^2}.
    \end{split}
\end{align}
The Pfaffian of a $2n\times 2n$ antisymmetric matrix $M$ is
\begin{align}
    \begin{split}
        \text{Pf}\left(M\right) &= \frac{1}{2^n n!}\sum_{\sigma}\text{sgn}\sigma \left[\prod_{i=1}^n M_{\sigma(2i-1),\sigma(2i)}\right]
    \end{split}
\end{align}
where $\sigma$ is a permutation of the matrix indices.
Expanding $\text{Pf}(\frac{1}{z_i-z_j})$ gives
\begin{align}
    \begin{split}
        \Psi_{\text{Pf}}(\{z_i\}) = \frac{1}{2^{N_e/2} (N_e/2)!}\sum_{\sigma}\text{sgn}\sigma \prod_{i=1}^{N_e /2}(z_{\sigma(2i-1)}-z_{\sigma(2i)})\prod_{j<(2i-1)}(z_{\sigma(2i-1)}-z_{\sigma(j)})^2e^{-\frac{1}{4\ell^2}\sum_i|z_i|^2}.
    \end{split}
\end{align}
We refer to each term in this sum over $\sigma$ as a ``Jastrow string". In each Jastrow string, each electron $i$ is paired with exactly one other electron $j$, meaning that the Jastrow string contains the factor $(z_i-z_j)$. For all other electrons $k\neq j$, there is an ``unpaired" quadratic Jastrow factor $(z_i-z_k)^2$.

The following three-body interaction produces the Pfaffian wavefunction as its exact zero-energy ground state when projected to the LLL \cite{rezayi2000incompressible}:
\begin{align}
\begin{split}
    \hat{V}_{\text{Pf}} &= \ell^{10}U_{\text{Pf}}\frac{1}{3!}\sum_{i}\sum_{j\neq i}\sum_{k\neq i,j}\nabla^4_i\nabla_j^2\delta^2(\bm{r}_i-\bm{r}_j)\delta^2(\bm{r}_j-\bm{r}_k).
\end{split}
\end{align}
To clarify, we note that an overall minus sign was incorrectly written in the definition of $\hat{V}_{\text{Pf}}$ in Ref. ~\cite{rezayi2000incompressible}. The factor $\ell^{10}U_{\text{Pf}}$, where $\ell$ is the magnetic length and $U_{\text{Pf}}$ has dimensions of energy, serves to give the operator $\hat{V}_{\text{Pf}}$ dimensions of energy.


We now prove that the Pfaffian parent Hamiltonian, $H_{\text{Pf}} = P_{\text{LLL}}\hat{V}_{\text{Pf}}P_{\text{LLL}}$ where $P_{\text{LLL}}$ projects onto the LLL, is positive semidefinite \cite{yoshioka2013quantum}. Let us decompose an arbitrary many-body wavefunction in the LLL $\Psi_{\text{LLL}}(\{z_i\})$ as
\begin{align}
    \begin{split}
        \Psi_{\text{LLL}} = pg
    \end{split}
\end{align}
where $p$ is a polynomial in $z_i$, $g=e^{-\frac{1}{4\ell^2}\sum_i|z_i|^2}$ in the symmetric gauge, and we leave the dependence on $\{z_i\}$ implicit. Due to Fermi statistics, $p \propto \prod_{i\neq j}(z_i - z_j)$. Now

\begin{align}
\begin{split}
    \bra{\Psi_{\text{LLL}}} \nabla_1^4 \nabla_2^2 \delta^2(\bm{r}_1-\bm{r}_2)\delta^2(\bm{r}_2-\bm{r}_3)\ket{\Psi_{\text{LLL}}}  &= \left(\prod_i \int \, d^2r_i\right)\left(\bar{\partial}^2_1\partial^2_1\bar{\partial}_2 \partial_2\delta^2(\bm{r}_1-\bm{r}_2)\delta^2(\bm{r}_2-\bm{r}_3)\right)(\bar{p}pg^2) \\
    &= \left(\prod_i \int \, d^2r_i\right)\delta^2(\bm{r}_1-\bm{r}_2)\delta^2(\bm{r}_2-\bm{r}_3) (\bar{\partial}^2_1\partial^2_1\bar{\partial}_2 \partial_2\bar{p}pg^2) \\
    &= \left(\prod_i \int \, d^2r_i\right)\delta^2(\bm{r}_1-\bm{r}_2)\delta^2(\bm{r}_2-\bm{r}_3) \bar{\partial}^2_1\partial^2_1\left[|\partial_2 p|^2g^2 + \bar{p}\bar{\partial}_2 \partial_2(pg^2) + p\bar{\partial}_2 \partial_2(\bar{p}g^2) \right] \\
    &= \left(\prod_{i\neq 3} \int \, d^2r_i\right)\delta^2(\bm{r}_1-\bm{r}_2) \bar{\partial}^2_1\partial^2_1\left(|\partial_2 p|^2g^2\right)|_{\bm{r}_3=\bm{r}_2} \\
    &= \left(\prod_{i\neq 3} \int \, d^2r_i\right)\delta^2(\bm{r}_1-\bm{r}_2)|\partial_1^2 \partial_2 p|^2g^2|_{\bm{r}_3=\bm{r}_2} \\
    & \geq 0
\end{split}
\end{align}
where we define $\partial_i = \partial_{x,i} -i\partial_{y,i}$ so that $\nabla^2_i = \bar{\partial}_i\partial_i$. In these manipulations, we repeatedly use the fact that if $p\propto (z_i-z_j)^n$, it must be acted on by $\partial_{i/j}^n$ to lose all factors of $(z_i-z_j)$ and thus not vanish when integrated against $\delta^2(\bm{r}_i-\bm{r}_j)$. We also use the fact that the part of $\partial_2 p |_{\bm{r}_3=\bm{r}_2}$ that survives the integral over $\bm{r}_3$ against $\delta^2(\bm{r}_2-\bm{r}_3)$ is $\propto (z_1-z_2)(z_1-z_3)|_{\bm{r}_3=\bm{r}_2}=(z_1-z_2)^2$ and thus needs to be acted on by $\partial_1^2$ to survive integration over $\bm{r}_2$ against $\delta^2(\bm{r}_1 - \bm{r}_2)$.

We now show that this integral vanishes for the Pfaffian wavefunction. We note that the only Jastrow string that survives in $|\partial_2p|_{\bm{r}_3=\bm{r}_2}$ is that in which particles $2$ and $3$ are paired. Because $(1,2)$ and $(1,3)$ are unpaired in this Jastrow string, $|\partial_2p|_{\bm{r}_3=\bm{r}_2}\propto (z_1-z_2)^2(z_1-z_3)^2|_{\bm{r}_3=\bm{r}_3} = (z_1-z_2)^4$. Thus, at least $4$ factors of $(z_1-z_2)$ remain in $|\partial_1^2 \partial_2 p|^2g^2|_{\bm{r}_3=\bm{r}_2}$ and the final line above vanishes.

\subsection{Equations for numerical calculation}

For the purpose of numerical calculation, it is useful to write the Pf parent interaction as
\begin{align}
\begin{split}\label{eq:VPfMomSpace}
    \hat{V}_{\text{Pf}} &= \frac{1}{3!}\frac{1}{A^2}\sum_{\bm{q}_1,\bm{q}_2,\bm{q}_3}v(\bm{q}_1,\bm{q}_2,\bm{q}_3):\rho(-\bm{q}_1)\rho(-\bm{q}_2)\rho(-\bm{q}_3):
\end{split}
\end{align}
where $\rho(\bm{q}) = \sum_i e^{-i\bm{q}\cdot\bm{r}_i}$ and
\begin{align}
    \begin{split}
        v(\bm{q}_1,\bm{q}_3,\bm{q}_3) &=-\ell^{10}U_{\text{Pf}} q_2^2 q_1^4 \delta_{\bm{0},\bm{q}_1+\bm{q}_2+\bm{q}_3}.
    \end{split}
\end{align}
The Pf parent Hamiltonian is then
\begin{align}
    \begin{split}
        H_{\text{Pf}} &= \frac{1}{3!}\sum_{\bm{k}_1,...,\bm{k}_6}V_{\bm{k}_1,\bm{k}_2,\bm{k}_3;\bm{k}_4,\bm{k}_5,\bm{k}_6}c^{\dag}_{\bm{k}_1}c^{\dag}_{\bm{k}_2}c^{\dag}_{\bm{k}_3}c_{\bm{k}_6}c_{\bm{k}_5}c_{\bm{k}_4}
    \end{split}
\end{align}
where
\begin{align}
    \begin{split}
    V_{\bm{k}_1,\bm{k}_2,\bm{k}_3;\bm{k}_4,\bm{k}_5,\bm{k}_6} &= \bra{0,\bm{k}_1;0,\bm{k}_2;0,\bm{k}_3}\hat{V}_{\text{Pf}} \ket{0,\bm{k}_4;0,\bm{k}_5;0,\bm{k}_6} \\
    &= \frac{1}{A^2}\sum_{\bm{q}_1,\bm{q}_2,\bm{q}_3}v(\bm{q}_1,\bm{q}_2,\bm{q}_3) \bra{0,\bm{k}_1}e^{i\bm{q}_1\cdot\bm{r}}\ket{0,\bm{k}_4}\bra{0,\bm{k}_2}e^{i\bm{q}_2\cdot\bm{r}}\ket{0,\bm{k}_5}\bra{0,\bm{k}_3}e^{i\bm{q}_3\cdot\bm{r}}\ket{0,\bm{k}_6}.
    \end{split}
\end{align}
The anti-Pfaffian (aPf) state is $\ket{\text{aPf}} = C_{\text{ph}}\ket{\text{Pf}}$ where we define the anti-unitary particle-hole conjugation operator $C_{\text{ph}}$ by the relation $C_{\text{ph}}\alpha c^{\dag}_{\bm{k}}C_{\text{ph}}^{-1} = \bar{\alpha}c_{\bm{k}}$, $C_{\text{ph}}\alpha c_{\bm{k}}C_{\text{ph}}^{-1} = \bar{\alpha}c^{\dag}_{\bm{k}}$ where $\alpha$ is a complex number, the overbar denotes complex conjugation, and $c^{\dag}_{\bm{k}}$ is a LLL creation operator. The aPf parent Hamiltonian is then 
\begin{align}
    \begin{split}
    H_{\text{aPf}} &= C_{\text{ph}}H_{\text{Pf}}C_{\text{ph}}^{-1} \\ &= \frac{1}{3!}\sum_{\bm{k}_1,...,\bm{k}_6}\bar{V}_{\bm{k}_1,\bm{k}_2,\bm{k}_3;\bm{k}_4,\bm{k}_5,\bm{k}_6} c_{\bm{k}_1}c_{\bm{k}_2}c_{\bm{k}_3}c^{\dag}_{\bm{k}_6}c^{\dag}_{\bm{k}_5}c^{\dag}_{\bm{k}_4}.
    \end{split}
\end{align}

Note that the definition of $C_{\text{ph}}$ implies $C_{\text{ph}}\ket{\text{empty band}} = \ket{\text{full band}}$ because $c_{\bm{k}}\ket{\text{empty band}}=0=C_{\text{ph}}c_{\bm{k}}\ket{\text{empty band}} =  C_{\text{ph}}c_{\bm{k}}C^{-1}_{\text{ph}}C_{\text{ph}}\ket{\text{empty band}} = c^{\dag}_{\bm{k}}\left(C_{\text{ph}}\ket{\text{empty band}}\right)$. Since the state $C_{\text{ph}}\ket{\text{empty band}}$ is annihilated by all $c^{\dag}_{\bm{k}}$, it is $\ket{\text{full band}}$.

We now discuss the particle-hole symmetry of LL-projected two-body interaction Hamiltonians. A generic band-projected Hamiltonian $H_{{\text{band}}}=P_{\text{band}}HP_{\text{band}}$ with a two-body interaction can be written as
\begin{align}
    \begin{split}
        H_{\text{band}} &= \sum_{\bm{k}}\varepsilon(\bm{k}) c^{\dag}_{\bm{k}}c_{\bm{k}} + \frac{1}{2}\sum_{\bm{k},\bm{k}',\bm{q}}V_{[\bm{k}+\bm{q}],[\bm{k}'-\bm{q}];\bm{k},\bm{k}'}c^{\dag}_{[\bm{k}+\bm{q}]}c^{\dag}_{[\bm{k}'-\bm{q}]}c_{\bm{k}'}c_{\bm{k}}.
    \end{split}
\end{align}
Its particle-hole conjugate is
\begin{align}
    \begin{split}
        C_{\text{ph}} H_{\text{band}} C_{\text{ph}}^{-1} &= \sum_{\bm{k}}\varepsilon(\bm{k})(1- c^{\dag}_{\bm{k}}c_{\bm{k}}) + \sum_{\bm{k}}\Sigma(\bm{k})\left( \frac{1}{2} - c^{\dag}_{\bm{k}}c_{\bm{k}}\right) + \frac{1}{2}\sum_{\bm{k},\bm{k}',\bm{q}}V_{[\bm{k}+\bm{q}],[\bm{k}'-\bm{q}];\bm{k},\bm{k}'}c^{\dag}_{[\bm{k}+\bm{q}]}c^{\dag}_{[\bm{k}'-\bm{q}]}c_{\bm{k}'}c_{\bm{k}}
    \end{split}
\end{align}
where $\Sigma(\bm{k}) = \sum_{\bm{k}'}\left(V_{\bm{k},\bm{k}';\bm{k},\bm{k}'} - V_{\bm{k},\bm{k}';\bm{k}',\bm{k}}\right)$ a Hartree-Fock self energy. In the $n\text{LL}$, both $\varepsilon(\bm{k})$ and $\Sigma(\bm{k})$ are $\bm{k}$-independent. This is a consequence of magnetic translation symmetry and can be verified by direct calculation using the results of section \ref{sec:LLtorus}. Upon restricting consideration to states for which $\nu = \frac{1}{2}$, the Hartree-Fock self-energy term vanishes and we find that $H_{n\text{LL}}$ is particle-hole symmetric: $C_{\text{ph}} H_{n\text{LL}} C_{\text{ph}}^{-1} = H_{n\text{LL}}$.

\section{Characterization of the $\nu = \frac{3}{2}$ states}

\subsection{Overlaps}



Because a generic Chern band lives in a non-LL Hilbert space (e.g. a torus in the continuum without a net magnetic flux or a lattice), the overlap between a many-body state in its Fock space and a many-body state in a LL Fock space 
is trivially zero. Therefore, direct comparison between many-body states in generic Chern band and LL Fock spaces requires a modified definition of overlap. A natural procedure is to map Fock states in the LL and Chern band Fock spaces to a new Fock space whose basis states are defined by sets $\{\bm{k}\}$ of occupied (magnetic) Bloch states: $\left(\prod_{\bm{k} \in \{\bm{k}\}} c^{\dag}_{\text{LL},\bm{k}}\right)\ket{\text{vac}} \rightarrow \ket{\{\bm{k}\}}$, $\left(\prod_{\bm{k} \in \{\bm{k}\}} c^{\dag}_{\text{band},\bm{k}}\right)\ket{\text{vac}} \rightarrow \ket{\{\bm{k}\}}$. Here, the vacuum state $\ket{\text{vac}}$ includes full bands below the band of interest.
Under this mapping, a Chern band many-body state $\ket{\Psi_{\text{band}}} = \sum_{\{\{\bm{k}\}\}}\alpha(\{\bm{k}\})\left(\prod_{\bm{k} \in \{\bm{k}\}} c^{\dag}_{\text{band},\bm{k}}\right)\ket{\text{vac}} \rightarrow \sum_{\{\{\bm{k}\}\}}\alpha(\{\bm{k}\})\ket{\{\bm{k}\}} \equiv \ket{\Psi'_{\text{band}}}$ and likewise a LL many-body state $\ket{\Psi_{\text{LL}}} = \sum_{\{\{\bm{k}\}\}}\beta(\{\bm{k}\})\left(\prod_{\bm{k} \in \{\bm{k}\}} c^{\dag}_{\text{LL},\bm{k}}\right)\ket{\text{vac}} \rightarrow \sum_{\{\{\bm{k}\}\}}\beta(\{\bm{k}\})\ket{\{\bm{k}\}} \equiv \ket{\Psi'_{\text{LL}}}$. The overlap is then $\mathcal{O} = |\braket{\Psi'_{\text{band}}}{\Psi'_{\text{LL}}}| = |\sum_{\{\{\bm{k}\}\}}\alpha^*(\{\bm{k}\}) \beta (\{\bm{k}\})|$. However, this procedure is physically ambiguous because $\mathcal{O}$ varies under the gauge transformation $c^{\dag}_{\text{band},\bm{k}} \rightarrow e^{i\theta(\bm{k})}c^{\dag}_{\text{band},\bm{k}}$. Nevertheless, several gauge fixing procedures have been shown to produce high overlaps between Chern band fractional Chern insulators and LL fractional quantum Hall states \cite{wu2012gauge, wu2013bloch}. 

In our case, working in a Landau level basis offers a practical advantage. Since the effective adiabatic Hamiltonian (Eq. 2 in the main text) is defined on a continuum torus with a net magnetic flux, each of its magnetic Bloch miniband eigenstates $\ket{i,\bm{k}}$ is a linear combination of LL magnetic Bloch states $\ket{n,\bm{k}}$ with different Landau Level indices at fixed $\bm{k}$: $\ket{i,\bm{k}} = \sum_n z_{in\bm{k}}\ket{n,\bm{k}}$ (see section \ref{sec:LLtorus}). We simply set the coefficient $z_{21\bm{k}}$ of the 1LL state in the second miniband state at each $\bm{k}$ to be real and positive. With this gauge choice, we define the overlap through the mapping described in the previous paragraph.

We define $\ket{\Psi_{i,\text{Pf}}}$ as the $i^{th}$ exact ground state of $H_{\text{Pf}}$. As stated in the main text and shown explicitly in Fig. \ref{fig:LLComparison}, there are six such ground states when the number of electrons is even and two when it is odd. Under the mapping described above, $\ket{\Psi_{i,\text{Pf}}}\rightarrow \ket{\Psi'_{i,\text{Pf}}}$. We note that there exists an interaction (defined in section \ref{section:furtherRemarks}) for every Landau level whose ground state upon projection at half filling becomes $\ket{\Psi'_{i,\text{Pf}}}$ under the mapping defined above. Therefore, we are free to think of $\ket{\Psi'_{i,\text{Pf}}}$ as originating from a LLL or 1LL (or $n$LL, for that matter) state as convenient. Analogous statements hold for the aPf.

The ground state overlap reported in the main text is defined as  $\mathcal{O} = \sqrt{\sum_{i\in \text{(a)Pf}} |\braket{\Psi'_{i,\text{(a)Pf}}}{\Psi'_{GS}}|^2}$ where $\ket{\Psi_{GS}}$ the exact ground state (not averaged over quasidegenerate ground state manifold) of the Hamiltonian of interest (in this case, the adiabatic model for $t$MoTe$_2$). Note that this definition of $\mathcal{O}$ is generalized from the one introduced earlier in this section to account for the degeneracy of the model ground states.

\subsection{Hamiltonian interpolation}

\begin{figure*}
    \centering
\includegraphics[width=\textwidth]{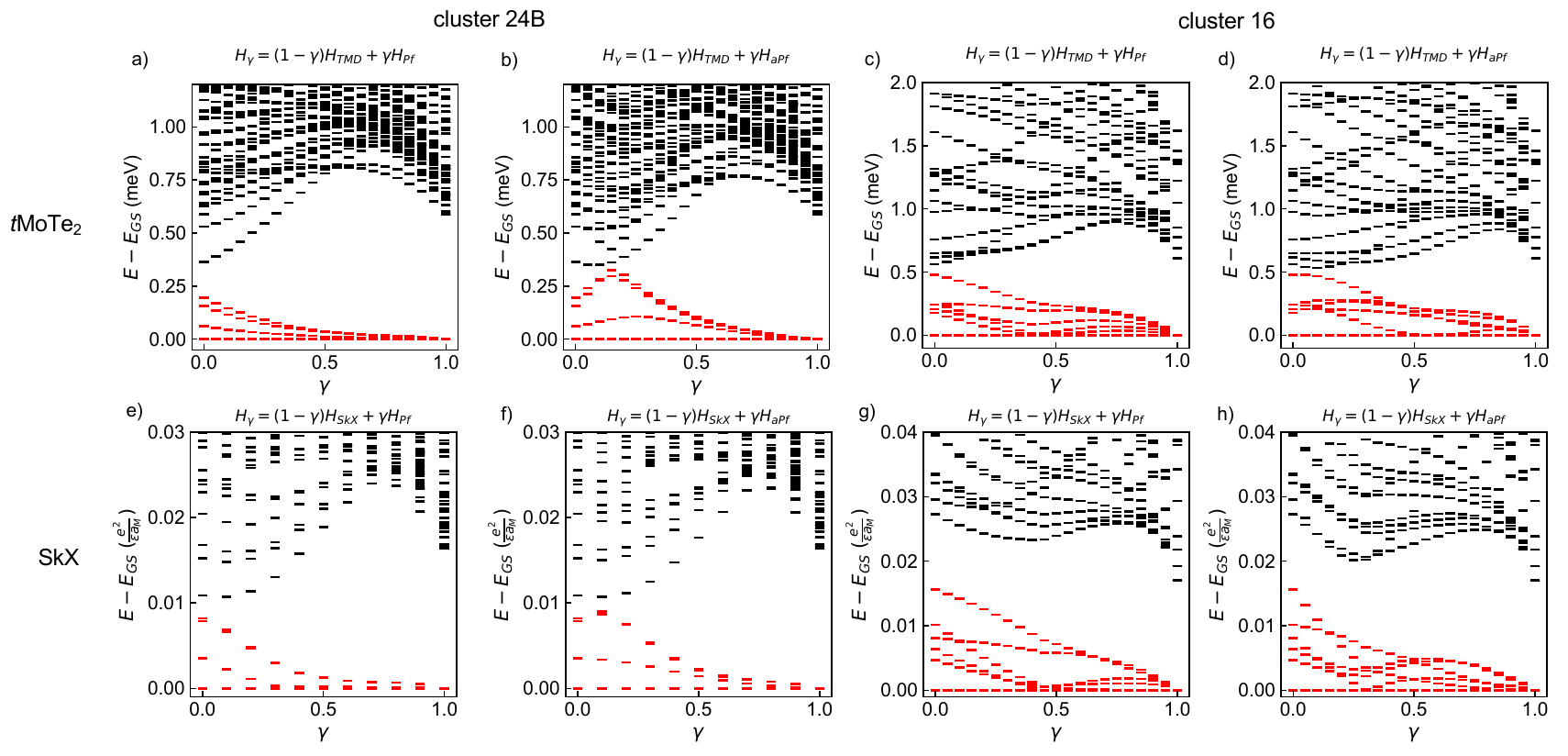}
    \caption{Energy spectrum of an interpolation Hamiltonian $H_{\gamma}$ interpolating between the adiabatic model for $t$MoTe$_2$ with a Coulomb interaction and the exact anti-Pfaffian (a,c) and anti-Pfaffian (b,d) three-body interaction Hamiltonians on clusters 16 (a,d) and 24B (c,d). (e-h) Analogous data for the SkX model. Parameters: (a-d) $\theta=2.5^{\circ}$, $\epsilon=5$; (e-h) $\alpha=1.0$, $N_0=0.28$, $\sqrt{\frac{4\pi}{\sqrt{3}}}\frac{e^2}{\epsilon a} = \frac{2\pi\hbar^2}{mA_{uc}}$.
    We set $U_{\text{Pf}} = \frac{e^2}{\epsilon \ell} = \sqrt{\frac{4\pi}{\sqrt{3}}}\frac{e^2}{\epsilon a_M}$.}
\label{fig:adConn}
\end{figure*}

In Fig. \ref{fig:adConn}, we show the energy spectra of Hamiltonians $H_{\gamma}$ interpolating between $H_{\text{TMD}}$ or $H_{\text{SkX}}$ and $H_{\text{Pf}}$ or $H_{\text{aPf}}$ on two example clusters. Precisely, the interpolation Hamiltonian is defined as

\begin{align}
    \begin{split}
        H_{\gamma} &= (1-\gamma) \left[ \sum_{\bm k} \tilde{\varepsilon}_{2}^{\text{band}}(\bm k)c^{\dag}_{2,\bm{k}}c_{2,\bm{k}} + \frac{1}{2}\sum_{\bm{k}_1,\bm{k}_2,\bm{k}_3,\bm{k}_4}V^{\text{band}}_{\bm{k}_1,\bm{k}_2;\bm{k}_3,\bm{k}_4}c^{\dag}_{2,\bm{k}_1}c^{\dag}_{2,\bm{k}_2}c_{2,\bm{k}_4}c_{2,\bm{k}_3} \right]  \\
        &+ \gamma \frac{1}{3!}\sum_{\bm{k}_1,...,\bm{k}_6}  V^{\text{Pf}}_{\bm{k}_1,\bm{k}_2,\bm{k}_3;\bm{k}_4,\bm{k}_5,\bm{k}_6}c^{\dag}_{2,\bm{k}_1}c^{\dag}_{2,\bm{k}_2}c^{\dag}_{2,\bm{k}_3}c_{2,\bm{k}_6}c_{2,\bm{k}_5}c_{2,\bm{k}_4}
    \end{split}
\end{align}
The superscript ``band" is either TMD or SkX. This expression assumes the same gauge choice that we use in our overlap definition in which $\braket{2,\bm{k}}{\text{1LL},\bm{k}} = z_{21\bm{k}}$ is real and positive. $c^{\dag}_{2,\bm{k}}$ creates a magnetic Bloch state $\ket{2,\bm{k}}$ in the second adiabatic $t$MoTe$_2$ or SkX miniband. $\tilde{\varepsilon}^{\text{band}}_{2}(\bm{k}) = \varepsilon^{\text{band}}_2(\bm{k}) + \Sigma^{\text{band}}_{22}(\bm{k})$ is the effective second miniband dispersion, including Hartree-Fock contributions from the full first band (see Sec. \ref{sec:HF}).

For the adiabatic model of $t$MoTe$_2$ (Fig. \ref{fig:adConn} (a-d)), the gap remains open throughout interpolation to the Pf parent Hamiltonian, but closes on cluster 24B and undergoes an anti-crossing on cluster 16 during interpolation to the aPf parent Hamiltonian. This contrasting behavior is consistent with our observation discussed in the main text that the Coulomb ground state has higher overlap with the Pf model states than it does with the aPf model states. It provides additional evidence that the Pfaffian is more likely to emerge in the thermodynamic limit than the anti-Pfaffian.

For the SkX model on cluster 16, the energy gap between the highest quasidegenerate ground state and next excited state remains open as a function of the interpolation parameter $\gamma$ upon interpolation to both the Pf and aPf. However, on cluster 24B, there appears to be a level anti-crossing during interpolation to $H_{\text{aPf}}$ while there is no such anticrossing throughout interpolation to $H_{\text{Pf}}$.

Upon interpolating between the Coulomb 1LL Hamiltonian and $H_{\text{Pf}}$ or $H_{\text{aPf}}$,  the gap does not close on all clusters we have tested (not shown). We refrain from interpreting the absence of gap closing in finite-size Hamiltonian interpolation as a definitive demonstration of ``adiabatic connectivity" because this interpretation would imply that the Pf and aPf states are adiabatically connected to each other. This cannot be true because the two model states correspond to distinct topological orders \cite{levin2007particle,lee2007particle} and distinct topological orders are not adiabatically connected by definition. The absence of gap closing during interpolation to both $H_{\text{Pf}}$ and $H_{\text{aPf}}$ is a finite-size artifact.


\subsection{Entanglement Spectroscopy}
Entanglement spectroscopy \cite{li2008entanglement} has been shown to be a powerful tool to probe possible topologically ordered states. Depending on the chosen cut, the low-lying entanglement states convey information about the underlying topological order. Here, we focus on particle entanglement spectrum (PES)\cite{sterdyniak2011extracting}. By partitioning the system in the particle space into two subspaces, $N = N_A + N_B$, PES is defined as the set $\{\xi = - \log \epsilon_i\}$ where $\{\epsilon_i\}$ are the eigenvalues of the reduced density matrix $\rho_A$ obtained by tracing $N_B$ particles, $\rho_A = \Tr_{N_B} \rho$ and $\rho = \sum_{i = 1}^{d} \ket{\Psi_i}\bra{\Psi_i}$ is the density matrix for $d$ quasi-degenerate ground states. 

For gapped topologically ordered phases, PES has been shown in numerous cases to exhibit an \textit{entanglement} gap, below which the number of states matches the dimension of the quasi-holes Hilbert space which is a unique characteristic of the underlying topological order. The counting rules can be derived in different ways \cite{bergholtz2008quantum,bernevig2008model}. The MR (Pfaffian or anti-Pfaffian) non-Abelian state satisfies the (2,4) counting rule \cite{ardonne2008degeneracy} which forbids no more than 2 particles in each 4 consecutive orbitals. 

\begin{figure*}
    \centering
\includegraphics[width=\textwidth]{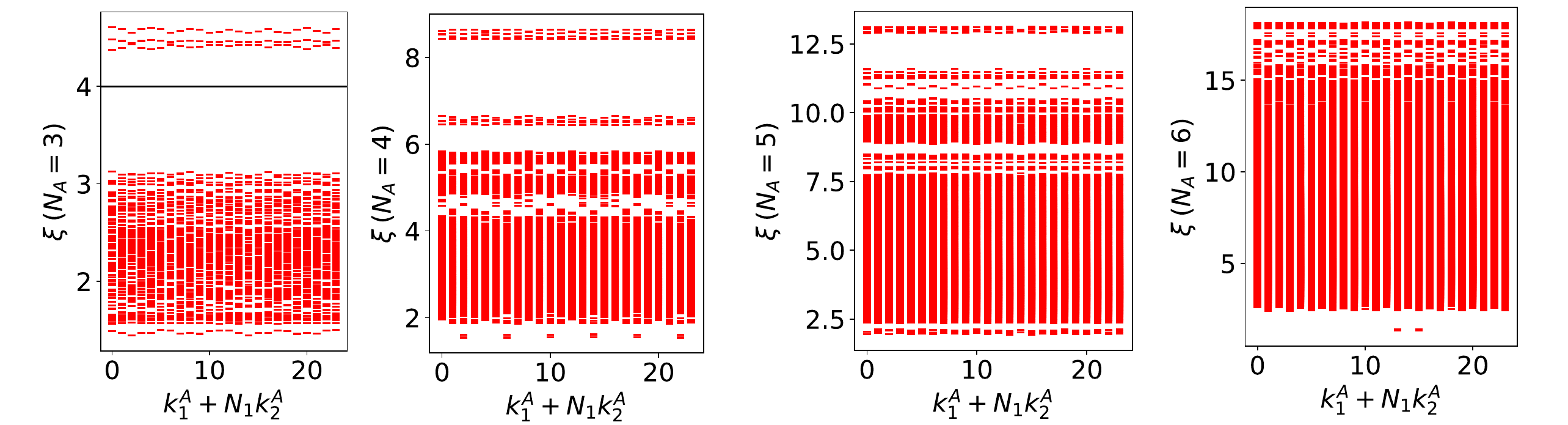}
    \caption{Particle entanglement spectrum (PES) for the SCB model calculated on cluster 24A  for different cuts obtained by tracing out \textit{electrons}. Out of all cuts considered, only the cut corresponding to leaving $N_A = 3$ particles (leftmost panel) leads to an entanglement gap with low-lying states consistent with the expected quasi-hole counting of the MR state. }
\label{fig:PES_skyrmion_electron}
\end{figure*}

\begin{figure*}
    \centering
\includegraphics[width=\textwidth]{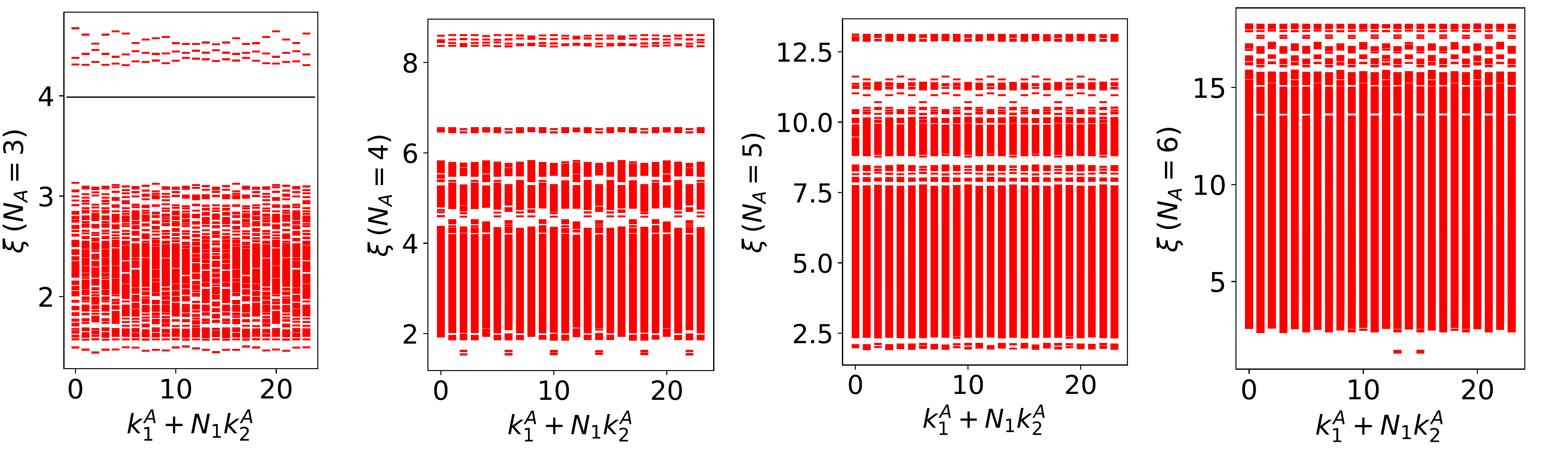}
    \caption{Particle entanglement spectrum (PES) for the SCB model calculated on cluster 24A  for different cuts obtained by tracing out \textit{holes}. Out of all cuts considered, only the cut corresponding to leaving $N_A = 3$ holes (leftmost panel) leads to an entanglement gap consistent with the expected quasi-hole counting of the MR state. }
\label{fig:PES_skyrmion_holes}
\end{figure*}

We calculate the PES for the SkX model by tracing either electrons (Fig. \ref{fig:PES_skyrmion_electron}) or holes (Fig. \ref{fig:PES_skyrmion_holes}). We trace out different number particles $N_B$ and analyze the resulting entanglement gaps. We only find a consistent entanglement gap with the MR non-Abelian state counting rule only for $N_A = 3$ particles (electrons or holes). While entanglement gaps exist for different $N_A$, we find that the number of states below these gaps doesn't match the (2,4) counting rule.  

\begin{figure*}
    \centering
\includegraphics[width=\textwidth]{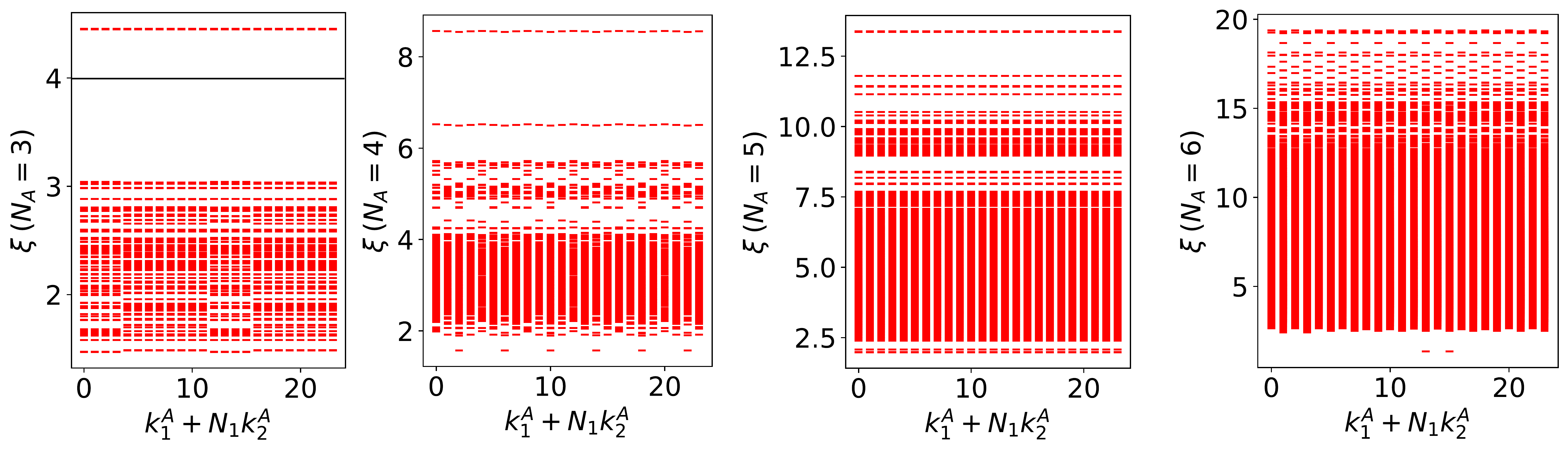}
    \caption{Particle entanglement spectrum (PES) for the 1LL on calculated on cluster 24A  for different cuts. Out of all cuts considered, only the cut corresponding to leaving $N_A = 3$ particles (holes) (leftmost panel) leads to an entanglement gap consistent with the expected quasi-hole counting of the MR state. }
\label{fig:PES_1LL}
\end{figure*}

We conjecture that this mismatch is due to the ground state being a superposition of the Pfaffian and the anti-Pfaffian with different weights. Indeed a similar behavior is observed in the 1LL (Fig \ref{fig:PES_1LL}. While the Pf and the aPf individually satisfy the (2,4) counting rule, it's not clear if a superposition of the two would satisfy the same counting rule. We leave this to future work. 

\subsection{Extended numerical data}

\begin{figure*}\label{fig:LLComparison}
    \centering
\includegraphics[width=\textwidth]{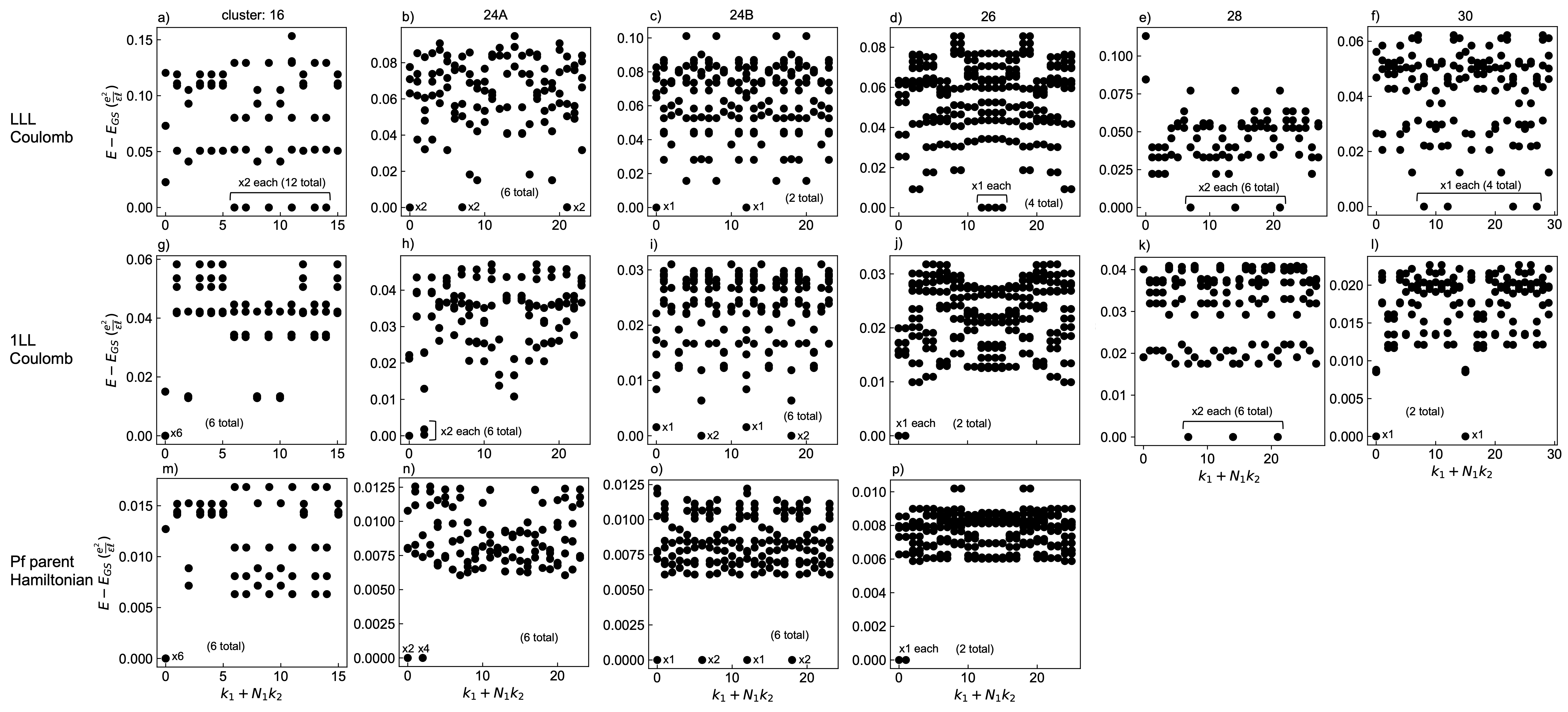}
    \caption{Comparison of low-energy spectra of LLL (a-f) and 1LL (g-l) at $\nu=\frac{1}{2}$ with Coulomb interaction and the exact Pf parent Hamiltonian (m-p), on several finite-size toruses. Ground states are labeled.  The Coulomb 1LL and exact Pf parent Hamiltonians' ground states' center-of-mass quasimomenta and degeneracies match in each case and differ from those of the Coulomb LLL Hamiltonian whose ground states correspond to a composite Fermi liquid phase. We set $U_{\text{Pf}} = \frac{e^2}{\epsilon \ell}$. The spectra of the aPf parent Hamiltonian (not shown) are identical to those of the Pf parent Hamiltonian.}
\end{figure*}

In Fig. \ref{fig:LLComparison}, we compare spectra of the Coulomb LLL and 1LL projected Hamiltonians with spectra of the Pf parent Hamiltonian on several finite-size clusters. These data show that the the center-of-mass momentum quantum numbers of composite Fermi liquid and Pf/aPf ground states are generally distinct. Therefore, the center-of-mass momentum quantum numbers of the ground states of a given Hamiltonian provide evidence to distinguish between these two candidate phases of matter.

\begin{figure*}
    \centering
\includegraphics[width=\textwidth]{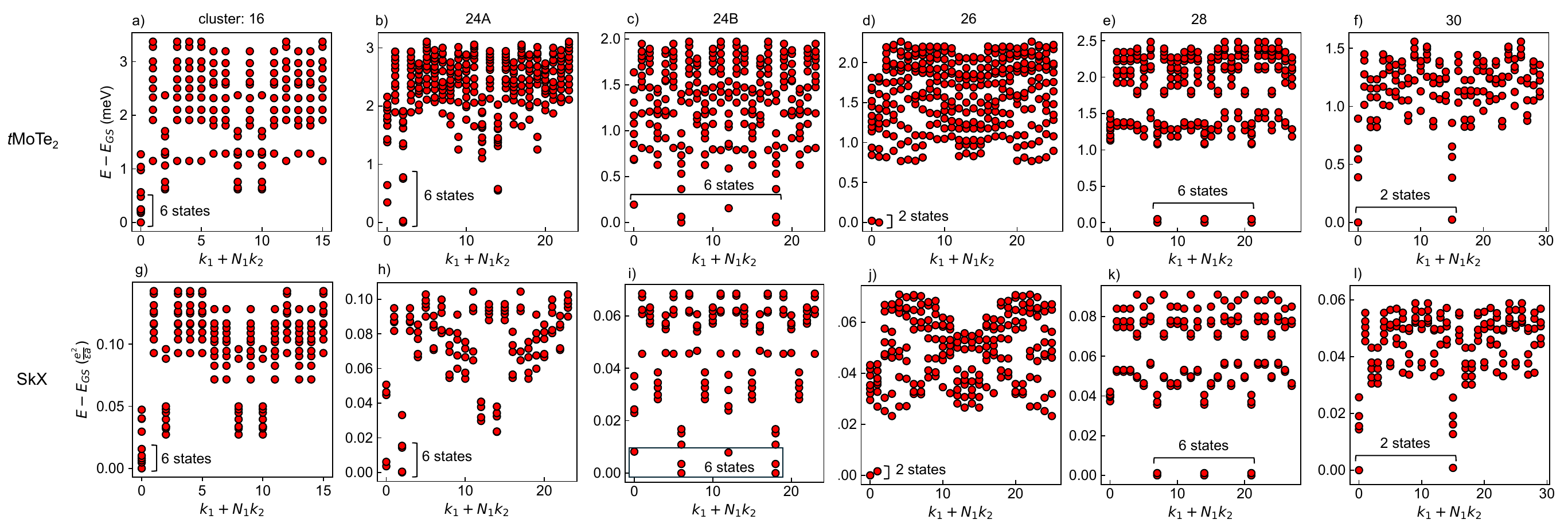}
    \caption{Additional many-body spectra of the adiabatic model for $t$MoTe$_2$ (a-f) and the SkX model (g-l) on several clusters. Parameters: (a-f) $\theta=2.5^{\circ}$, $\epsilon=5$; (g-l) $\alpha=1.0$, $N_0=0.28$, $\sqrt{\frac{4\pi}{\sqrt{3}}}\frac{e^2}{\epsilon a} = \frac{2\pi\hbar^2}{mA_{uc}}$.}
\label{fig:exampleSpectra}
\end{figure*}

In Fig. \ref{fig:exampleSpectra} we show many-body spectra for the SkX and adiabatic $t$MoTe$_2$ models on additional clusters. In all cases except for the $t$MoTe$_2$ adiabatic model on cluster $24$A, the ground state momentum sectors and degeneracies are consistent with Pfaffian or antiPfaffian states. We attribute the extraneous state lower in energy than some one the six putative quasidegenerate ground states on cluster $24$A for the $t$MoTe$_2$ adiabatic model to finite size effects because the expected quasi-degenerate ground states are the lowest energy states on all other clusters. Further, upon increasing interaction strength, this extraneous state moves higher in energy than the six putative non-Abelian ground states (not shown).

\begin{figure*}
    \centering
\includegraphics[width=0.3\textwidth]{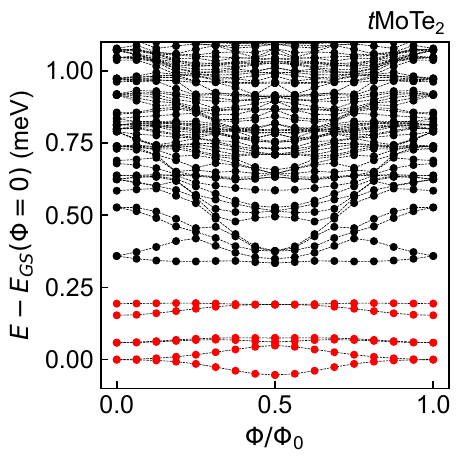}
    \caption{Spectral flow of the $t$MoTe$_2$ adiabatic model under flux insertion through a handle of the torus. Cluster 24B is used and the six quasi-degenerate ground states are highlighted in red. $\Phi$ is the inserted magnetic flux and $\Phi_0 = hc/e$ is the flux quantum. Parameters: $\theta=2.5^{\circ}$, $\epsilon=5$.}
\label{fig:tMoTe2FluxThreading}
\end{figure*}

In Fig. \ref{fig:tMoTe2FluxThreading} we show spectral flow of the adiabatic model of $t$MoTe$_2$ under flux insertion through a handle of the torus. The energy gap between the six quasi-degenerate ground states and the next excited state remains open as flux is inserted, suggesting that the energy gap is a genuine feature of the thermodynamic limit rather than a finite-size artifact.

\begin{figure*}
    \centering
\includegraphics[width=0.4\textwidth]{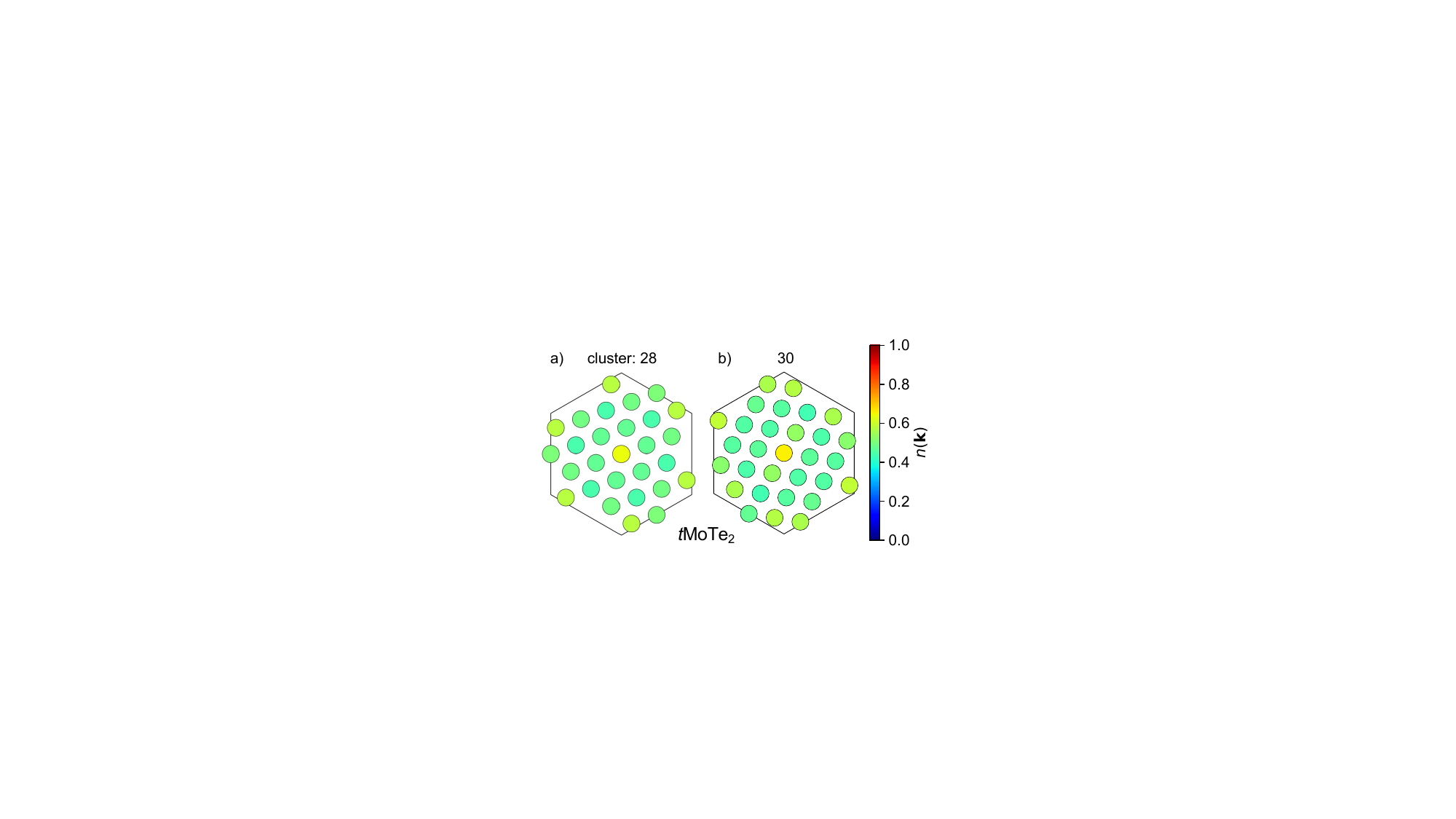}
    \caption{Example momentum distribution functions $n(\bm{k})= \frac{1}{N_{GS}}\sum_{i \in GS} \bra{i}c^{\dag}_{2,\bm{k}}c_{2,\bm{k}}\ket{i}$ where $c^{\dag}_{2,\bm{k}}$ creates a particle in the magnetic Bloch state with magnetic crystal momentum $\bm{k}$ in the second miniband of the $t$MoTe$_2$ adiabatic model. GS is the set of exactly degenerate lowest-energy many-body states. Parameters: $\theta=2.5^{\circ}$, $\epsilon=5$.}
\label{fig:tmdThreeHalvesOcc}
\end{figure*}

In Fig. \ref{fig:tmdThreeHalvesOcc} we show the momentum distribution functions $n(\bm{k})$ in the exact ground state of the adiabatic $t$MoTe$_2$ model on several finite clusters. $n(\bm{k})$ is smooth and nearly uniform.

\begin{figure*}
    \centering
\includegraphics[width=0.8\textwidth]{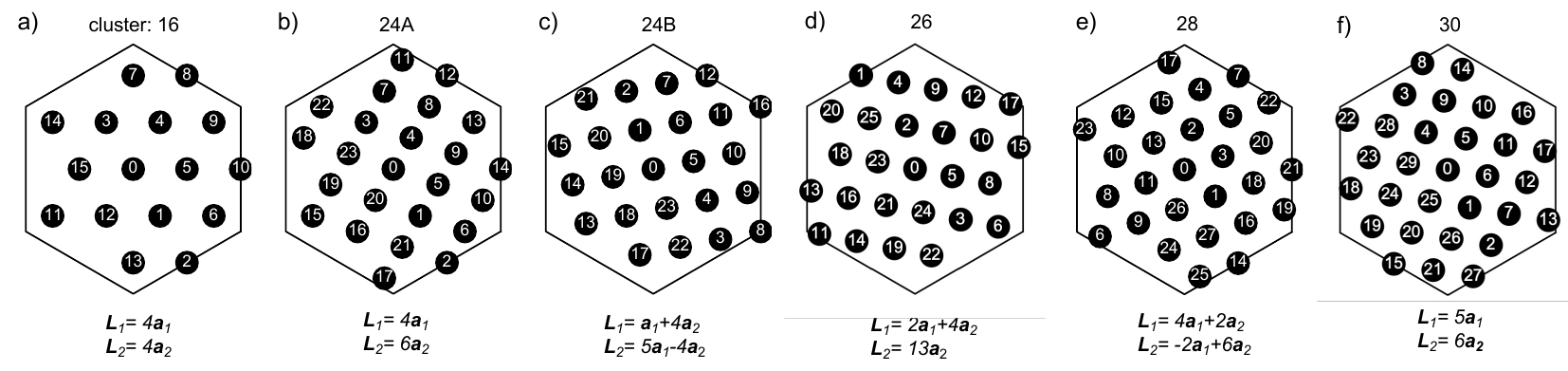}
    \caption{Momentum space mesh diagrams of the finite-size clusters used in this work. $\bm{L}_i$ are the torus boundary vectors. $\bm{a}_{1} = a_{m} (\frac{\sqrt{3}}{2},-\frac{1}{2})$ and $\bm{a}_{2} = a_{m} (\frac{\sqrt{3}}{2},+\frac{1}{2})$. $\bm T_i = \frac{2\pi \epsilon_{ij}\bm L_j \times \hat{z}}{|\bm L_1 \times \bm L_2|}$ are the momentum basis vectors. Mesh points $\bm{k}=k_1\bm{T}_1+k_2\bm{T}_2$ (modulo reciprocal lattice vector) points are labeled by the integer $k=k_1+ N_1 k_2$. $k_i$ takes on all integer values from $0$ through $(N_i-1)$.}
\label{fig:clusters}
\end{figure*}

Finally, in Fig. \ref{fig:clusters}, we define the finite-size systems used in this work and show their corresponding (magnetic) Brillouin zone meshes.

\subsection{Further remarks}\label{section:furtherRemarks}
Given that $\hat{{V}}_\text{Pf}$ has zero-energy Pfaffian ground states upon projection to the LLL, one might expect that it also has ground states in the Pfaffian phase upon projection to the 1LL. We find by direct calculation that this is not the case. In particular, $P_{1\text{LL}}\hat{V}_{\text{Pf}}P_{1\text{LL}}$ produces ground states with different center of mass magnetic momentum quantum numbers and degeneracies than does $H_{\text{Pf}} = P_{0\text{LL}}\hat{V}_{\text{Pf}}P_{0\text{LL}}$. Similarly, projecting $\hat{V}_{\text{Pf}}$ to either the second SkX or adiabatic TMD bands does not produce a spectrum consistent with a Pf ground state.

Formally, we can define a three-body interaction $\hat{V}_{\text{Pf}, n\text{LL}}$ such that the spectrum of $P_{n\text{LL}}\hat{V}_{\text{Pf}, n\text{LL}}P_{n\text{LL}}$ is independent of $n$ by replacing $v(\bm{q}_1,\bm{q}_3,\bm{q}_3)$ in Eq. \ref{eq:VPfMomSpace} with $v_{n\text{LL}}(\bm{q}_1,\bm{q}_3,\bm{q}_3) = -\frac{\ell^{10} U_{\text{Pf}} q_1^4q_2^2}{L_n(\ell^2 q_1^2/2)L_n(\ell^2 q_2^2/2)L_n(\ell^2 q_3^2/2)}\delta_{\bm{q}_1+\bm{q}_2+\bm{q}_3,\bm{0}}$. The Laguerre polynomials in the denominator serve to cancel out factors in the $n\text{LL}$ form factors not present in the LLL from factor (see section \ref{sec:LLtorus}). We find that projecting $\hat{V}_{\text{Pf}, 1\text{LL}}$ to the second SkX or adiabatic $t$MoTe$_2$ band does not produce a Pf-like spectrum.

Finally, we remark that we have not found evidence for a non-Abelian state at $n=\frac{3}{2}$ using the continuum model parameters of Ref. \cite{reddy2023fractional} without making the adiabatic approximation.


\section{Working on a torus with a net magnetic flux}\label{sec:LLtorus}

Here we describe how we perform calculations on the adiabatic effective Hamiltonian $H_{\text{eff}}$ (Eq. 2 in the main text), which contains a net magnetic flux, on a torus. We take an algebraic approach, exploiting the magnetic translation algebra. Our discussion builds on the Supplementary Material of Ref. \cite{wang2021exact}.

We begin with the standard LL Hamiltonian, $H_{\text{LL}}=\bm{\pi}^2/(2m)=\hbar\omega_c(a^{\dag}a+1/2)$. Here $\bm{\pi} = \bm{p}+\frac{e}{c}\bm{A}(\bm{r})$, $\bm{B}(\bm{r}) = \nabla \times \bm{A}(\bm{r}) =-B_0\hat{z}$ (this sign choice, with $B_0>0$ and (electron charge)$=-e<0$, gives holomorphic rather than antiholomorphic LLL wavefunctions, up to the Gaussian factor), $a=\frac{\ell}{\hbar \sqrt{2}}(\pi_x+i\pi_y)$, $\omega_c = eB_0/(mc)$, $2\pi\ell^2B_0=\Phi_0$, and $\Phi_0=hc/e$ is the flux quantum.
$H_{\text{LL}}$ is not translation invariant: $[T(\bm{d}),H_{\text{LL}}]\neq 0$ where $T(\bm{d})=e^{i\bm{p}\cdot\bm{r}/\hbar}$ is the translation operator. However, since $\bm{B}(\bm{r}) = \nabla\times\bm{A}(\bm{r})$ is translation invariant, $\bm{\pi}$ is translation invariant \emph{up to a gauge transformation}: $T(\bm{d})\bm{\pi} T^{\dag}(\bm{d}) = \bm{\pi} + \nabla\chi_{\bm{d}}(\bm{r}) = e^{-i\chi_{\bm{d}}(\bm{r})}\bm{\pi}e^{i\chi_{\bm{d}}(\bm{r})}$ where $\chi_{\bm{d}}(\bm{r}) = \frac{2\pi}{\Phi_0}\int_{\bm{r}}^{\bm{r}+\bm{d}}\,d\bm{r}' \cdot \bm{A}(\bm{r}')$. This motivates the definition of a magnetic translation operator \cite{zak1964magnetic,girvin2019modern}
\begin{align}
    \begin{split}
        t(\bm{d}) &= e^{i\chi_{\bm{d}}(\bm{r})}T(\bm{d})
    \end{split}
\end{align}
that bakes the gauge transformation into the translation operator and is a symmetry of the Hamiltonian: $[t(\bm{d}),\bm{\pi}]=0=[t(\bm{d}),H_{\text{LL}}]$.

It follows from the definition of $\chi_{\bm{d}}(\bm{r})$ that
\begin{align}\label{eq:MTA}
    \begin{split}
        t(\bm{d}')t(\bm{d}) &= e^{i\frac{1}{2\ell^2}\bm{d}'\wedge \bm{d}}t(\bm{d}+\bm{d}')
    \end{split}
\end{align}
where $\bm{d}'\wedge\bm{d}=(\bm{d}'\times\bm{d})\cdot\hat{z}$. Eq. \ref{eq:MTA} is the magnetic translation algebra. Note that $\bm{d}'\wedge\bm{d}/\ell^2 = 2\pi\Phi(\bm{d},\bm{d}')/\Phi_0$ where $\Phi(\bm{d},\bm{d}')$ is the magnetic flux enclosed by the rhombus with sides $\bm{d},\bm{d}'$.

It is useful to decompose the position operator into a ``Landau orbit" $\tilde{\bm{R}} = \frac{\ell^2}{\hbar}\hat{z}\times \bm{\pi}$ and ``guiding center" $\bm{R}$: $\bm{r}=\tilde{\bm{R}}+\bm{R}$ \cite{wang2021exact,haldane2018origin}. The cartesian components of the guiding center and Landau orbits obey the algebra
\begin{align}
    \begin{split}
        [R_i,R_j] = -i\epsilon_{ij}\ell^2 ;\; [\tilde{R}_i,\tilde{R}_j] = +i\epsilon_{ij}\ell^2; \; [R_i,\tilde{R}_j] =0.
    \end{split}
\end{align}
The Landau level index lowering operator is related to the Landau orbit as $a = \frac{1}{\ell \sqrt{2}}(\tilde{R}_y -i\tilde{R}_x)$.

In any gauge in which $\bm A(\bm{r})+ \bm  A(\bm{r}') = \bm A(\bm{r}+\bm{r}')$, the magnetic translation operator becomes
\begin{align}
    \begin{split}
        t(\bm{d}) &= e^{i\bm{P}\cdot\bm{d}/\hbar}
    \end{split}
\end{align}
where $\bm{P} = \frac{\hbar}{\ell^2}\hat{z}\times\bm{R}$ is the generator of magnetic translations. It is useful to define the magnetic momentum boost operator
\begin{align}
\begin{split}
    \tau(\bm{q}) &= e^{i\bm{q}\cdot\bm{R}}
\end{split}
\end{align}
which is equivalent to the magnetic translation operator: $t(\bm{d}) = \tau(\bm{d}\times \hat{z}/\ell^2)$. Written in terms of the magnetic momentum boost operators, the magnetic translation algebra is
\begin{align}
    \begin{split}
        \tau(\bm{q}')\tau(\bm{q}) &= e^{i\frac{\ell^2}{2}\bm{q}'\wedge\bm{q}}\tau(\bm{q}'+\bm{q}).
    \end{split}
\end{align}
Yet another useful way to express the magnetic translation algebra is
\begin{align}\label{eq:mixedMagTrans}
    \begin{split}
        t(\bm{d})\tau(\bm{q}) = e^{i\bm{q}\cdot\bm{d}}\tau(\bm{q})t(\bm{d})
    \end{split}.
\end{align}

We work on a torus defined by the two primitive boundary vectors $\bm{L}_1$, $\bm{L}_2$. A set of magnetic boundary conditions $\Phi_i$ defines a single-particle Hilbert space in which all states $\ket{\psi}$ obey the equation $t(\bm{L}_i)\ket{\psi}=e^{i2\pi\Phi_i}\ket{\psi}$ \cite{haldane1985periodic,haldane1985many}. The number of flux quanta through the surface of the torus is $N_{\Phi} = B_0 \bm{L}_1\wedge\bm{L}_2 /(2\pi\ell^2)$. The magnetic boundary condition requires $[t(\bm{L}_1),t(\bm{L}_2)]=0$ which implies that $N_{\Phi}$ must be an integer. We define the primitive wavevectors $\bm{Q}_1, \bm{Q}_2$ by the condition that $\tau(N_{\Phi}\bm{Q}_i) = t(\bm{L}_i)$, which gives $\bm{Q}_i = \epsilon_{ij} \bm{L}_j\times\hat{z}/(N_{\Phi}\ell^2)$. For $\tau(\bm{q})\ket{\psi}$ to belong to the same Hilbert space as $\ket{\psi}$ (that is, obey the same boundary condition), it must be that $[\tau(\bm{q}),\tau(\bm{L}_i)] = [\tau(\bm{q}),\tau(N_{\Phi}\bm{Q}_i)] =0$. In turn, this requires $\bm{q} = q_1\bm{Q}_1 + q_2\bm{Q}_2$ for integer $q_1$, $q_2$. These $\bm{q}$ are the allowed (=boundary-condition-preserving) wavevectors.

We define primitive magnetic lattice vectors $\bm{a}_i$ that obey condition $\bm{a}_1\wedge\bm{a}_2 = 2\pi\ell^2$ -- that is, that the magnetic unit cell, the rhombus defined by $\bm{a}_1$ and $\bm{a}_2$, encloses one flux quantum. Additionally, we require that the magnetic unit cell tile the torus or, equivalently, that there exist a matrix $M$ with integer elements such that $\bm{L}_i = M_{ij}\bm{a}_j$. It follows from the magnetic translation algebra that the set of operators $t(\bm{A})$ mutually commute where $\bm{A}=A_1\bm{a}_1+A_2\bm{a}_2$ (with integer $A_i$) is a magnetic lattice vector. Because $[H_{\text{LL}},t(\bm{A})]=0$, we can define a LL magnetic Bloch basis $\ket{n,\bm{k}}$ that simultaneously diagonalizes $H_{\text{LL}}$ and $t(\bm{A})$. Here $n$ is the LL index and $\bm{k}$ labels the eigenvalues of $t(\bm{A})$ in a way we now discuss.

Let $\ket{0,\bm{0}}$ be a LLL magnetic Bloch state and define $\phi_i$ by the equation $t(\bm{a}_i)\ket{0,\bm{0}} = e^{i2\pi\phi_i}\ket{0,\bm{0}}$. We note that the magnetic primitive lattice vector boundary condition $\phi_i$ is related to the torus magnetic boundary condition $\Phi_i$ as $\Phi_i = M_{ij}\phi_j +\frac{1}{2}M_{i1}M_{i2}$. Given $\ket{0,\bm{0}}$, the general LL magnetic Bloch state is
\begin{align}
    \begin{split}
        \ket{n,\bm{k}} = \ket{n} \otimes \ket{\bm{k}}
    \end{split}
\end{align}
where 
\begin{align}
    \begin{split}
        \ket{n} &= \frac{(a^{\dag})^n}{\sqrt{n!}}\ket{0}
    \end{split}
\end{align}
and
\begin{align}
    \begin{split}
        \ket{\bm{k}} &= \tau(\bm{k})\ket{\bm{0}}
    \end{split}
\end{align}
where $\bm{k}$ is an allowed wavevector as defined above.

With these definitions, the eigenvalue equation
\begin{align}
    \begin{split}
     t(\bm{a}_i)\ket{\bm{k}} = e^{i(\bm{k}\cdot\bm{a}_i + 2\pi\phi_i)}\ket{\bm{k}}   
    \end{split}
\end{align}
follows from the magnetic translation algebra, Eq. \ref{eq:mixedMagTrans}. We emphasize that it is not possible to satisfy the equation $t(\bm{A})\ket{\bm{k}}=e^{i(\bm{k}\cdot\bm{A} + 2\pi\phi_i)}\ket{\bm{k}}$ (analogous to the ordinary Bloch case) for all magnetic lattice vectors $\bm{A}$ because it contradicts the magnetic translation algebra. 

We define the primitive magnetic reciprocal lattice vectors $\bm{b}_i = \epsilon_{ij} \bm{a}_j\times \hat{z}/ \ell^2$ to obey the usual $\bm{a}_i\cdot\bm{b}_j = 2\pi\delta_{ij}$. From this definition, we have $t(\bm{a}_i) = \tau(-\epsilon_{ij}\bm{b}_j)$. Since $\ket{\bm{k}}$ is an eigenstate of $t(\bm{A})$, it must also be an eigenstate of all $\tau(\bm{g})$ where $\bm{g}=g_1\bm{b}_1+g_2\bm{b}_2$ for integer $g_1,g_2$ is a magnetic reciprocal lattice vector. Therefore, $\ket{\bm{k}}$ and $\ket{\bm{k}+\bm{g}}$ are redundant labels for the same quantum state differing only by an overall phase. The number of unique states $\ket{\bm{k}}$ is therefore equal to the number of allowed $\bm{q}$ that are unique modulo $\bm{g}$. This number is $N_{\Phi}$, the expected LL degeneracy.

We now write the wavefunction of the state $\ket{0,\bm{0}}$ only to show that it exists. We will not use it for calculations. In the symmetric gauge $\bm{A}(\bm{r}) = \frac{\hbar}{2\ell^2}\bm{r}\times\hat{z}$, the wavefunction of this state is
\begin{align}
    \begin{split}
        \braket{z,\bar{z}}{0,\bm{0}} &= \tilde{\sigma}\left(z+\frac{\phi_i-\frac{1}{2}}{\bar{a}_i}\right)e^{-|z|^2/2}
    \end{split}
\end{align}
where $z=\frac{1}{\ell \sqrt{2}}(x+iy)$ and $\tilde{\sigma}(z)$ is the modified Weierstrass sigma function \cite{haldane2018modular,haldane2018origin,wang2021exact}, which is holomorphic (entire) and quasiperiodic with respect to the magnetic lattice (that is, $\{\bm A\}$ sent to the complex plane). The property $t(\bm{a}_i)\ket{0,\bm{0}} = e^{i2\pi\phi_i}\ket{0,\bm{0}}$ can be verified by using the explicit symmetric gauge form of $t(\bm{a}_i)$ and the quasiperiodicity of the modified Weierstrauss sigma function under magnetic lattice translations: $\tilde{\sigma}(z+a_i) = - e^{\bar{a}_i(z+\frac{a_i}{2})}\tilde{\sigma}(z)$ \cite{wang2021exact}. This wavefunction lives in the LLL because it is the product of a Gaussian and a holomorphic function and thus is annihilated by the LL index lowering operator $a$.

The central ingredient in our numerical calculations is the matrix element
\begin{align}
    \begin{split}
        \bra{m,\bm{p}}e^{i\bm{q}\cdot\bm{r}}\ket{n,\bm{k}} &= \bra{m}e^{i\bm{q}\cdot\tilde{\bm{R}}}\ket{n}\bra{\bm{p}}\tau(\bm{q})\ket{\bm{k}}.
    \end{split}
\end{align}
We have
\begin{align}
    \begin{split}
        \bra{m}e^{i\bm{q}\cdot\tilde{\bm{R}}}\ket{n} &= e^{-\ell^2|\bm{q}|^2/4}G_{mn}\left(q\right)
    \end{split}
\end{align}
where $q= \ell(q_x+iq_y)/\sqrt{2}$ and

\begin{align}
\begin{split}
        G_{mn}(f) &= \bra{m}e^{fa^\dag}e^{-\bar{f} a}\ket{n} \\
        &= \begin{cases}
        f^{m-n}\sqrt{\frac{n!}{m!}}L_{n}^{m-n}(|f|^2) & n\leq m \\
        (-\bar{f})^{n-m}\sqrt{\frac{m!}{n!}}L_{m}^{n-m}(|f|^2) & n>m.
    \end{cases}
\end{split}
\end{align}
This can be derived by noting that $i\bm{q}\cdot\tilde{\bm{R}} = (q a^{\dag} - \bar{q} a)$, using the BCH formula to show $e^{i\bm{q}\cdot\tilde{\bm{R}}} = e^{(q a^{\dag} - \bar{q} a)} = e^{-|q|^2/2}e^{qa^{\dag}}e^{-\bar{q}a}$, Taylor expanding these exponentials, simplifying using harmonic oscillator algebra, and comparing to the power series definition of the generalized Laguerre polynomials $L^{\alpha}_{n}(x)$. We caution against potential numerical instability in computing the generalized Laguerre polynomials for large LL index. We recommend using the three-term recurrence relation rather than the closed-form power series formula to evaluate the generalized Laguerre polynomials.

Next, we have
\begin{align}
    \bra{\bm{p}}\tau(\bm{q})\ket{\bm{k}} &= e^{i\frac{\ell^2}{2}\left(-\bm{p}\wedge(\bm{k}+\bm{q}) +\bm{q}\wedge\bm{k}\right)}\sum_{\bm{g}}\delta_{\bm{k}+\bm{q}-\bm{p},\bm{g}}\eta(\bm{g})
\end{align}
where
\begin{align}
    \begin{split}
        \eta(\bm{g}) &= \bra{\bm{0}}\tau(\bm{g})\ket{\bm{0}}.
    \end{split}
\end{align}
This follows from the definition of $\ket{\bm{k}}$ and the magnetic translation algebra. Setting $\phi_1=\phi_2=\frac{1}{2}$ gives $\eta(\bm{g}) = +1$ if both $g_1$ and $g_2$ are even and $-1$ otherwise. With this choice of $\phi_i$, the wavefunction $\braket{z,\bar{z}}{0,\bm{0}}$ has its zero at the origin and the collective set of zeros of $\braket{z,\bar{z}}{0,\bm{k}}$ over all $\bm{k}$ is inversion symmetric about the origin. Therefore, it best respects the point group symmetries about the origin of the Hamiltonian at hand. For this reason, it corresponds most directly to periodic boundary conditions in the absence of magnetic flux and is the choice we use throughout this work.

The Hamiltonian for an electron in a periodic scalar potential and magnetic field with net one flux quantum per unit cell is
\begin{align}
    \begin{split}\label{eq:HPeriodicB}
        H &= \frac{\bm{\pi}^2}{2m} + V(\bm{r}) \\
        &= \hbar\omega_c\left[a^{\dag}a + \frac{1}{2} + \frac{\delta A_+ \delta A_- }{(\ell B_0)^2} +\frac{\delta A_-}{\ell B_0}a +a^{\dag}\frac{\delta A_+}{\ell B_0} + \frac{1}{2}\frac{\delta B}{B_0} + \frac{V}{\hbar\omega_c}\right]
    \end{split}
\end{align}
where $\omega_c = e|B_0|/(mc)$ is the cyclotron frequency, $B_0$ is the average magnetic field, and we leave the $\bm{r}$-dependence implicit in the second line. The effective Hamiltonian for SCB models, Eq. 2 in the main text, takes this form. Here we define $\bm{B}(\bm{r}) = -(B_0 + \delta B)\hat{z}$ where $\delta B$ has zero spatial average. Similarly, we define $\bm{A} = \bm{A}_0 + \delta\bm{A}$ such that $\nabla\times\bm{A}_0 = -B_0\hat{z}$ and $\nabla\times\delta\bm{A} = -\delta B \hat{z}$. Give a Fourier expansion of the magnetic field fluctuations $\delta B(\bm{r})=\sum_{\bm{g}}\delta B_{\bm{g}}e^{i\bm{g}\cdot\bm{r}}\hat{z}$, the corresponding vector potential is $\delta\bm{A} = i\sum_{\bm{g}}(g_y, -g_x)\frac{(-\delta B_{\bm{g}})}{|\bm{g}|^2}e^{i\bm{g}\cdot\bm{r}}\equiv \sum_{\bm{g}}\delta\bm{A}_{\bm{g}}e^{i\bm{g}\cdot\bm{r}}$. Also, $\delta A_{\pm} = (\delta A_x \pm i\delta A_y)/\sqrt{2}$. Since $\delta A(\bm{r})$ is periodic, the Bravais lattice magnetic translation operators $t(\bm{A})$ are unmodified from the uniform-field case discussed above.

In practice, we calculate the effective magnetic field and scalar potential Fourier coefficients of an SCB model by numerically evaluating the spin texture $\bm{S}(\bm{r})$ on a fine, uniform grid spanning a single Bravais lattice unit cell, calculating derivatives entering the effective magnetic field and the $(\partial_i\hat{\bm{S}})^2$ term by finite difference, and then numerically Fourier transforming.

The formulas we provide here allow for numerical diagonalization of Eq. \ref{eq:HPeriodicB} in the magnetic Bloch basis $\ket{n,\bm{k}}$. The resultant magnetic Bloch band eigenstates $\ket{i,\bm{k}}$ where $i$ is a band index are linear combinations of LL magnetic Bloch states at the same $\ket{\bm{k}}$: $\ket{i,\bm{k}} = \sum_n z_{in\bm{k}}\ket{n,\bm{k}}$. The LL weight from the main text is then $W_i = \frac{1}{N_{uc}}\sum_{\bm{k}}|z_{i(i-1)\bm{k}}|^2$. Interaction matrix elements can be calculated from $\bra{m,\bm{p}}e^{-i\bm{q}\cdot\bm{r}}\ket{n,\bm{k}}$ as usual. Explicitly, the LL basis matrix elements of a two-body interaction $\hat{V}$ with Fourier transformed interaction potential $v(\bm{q})$ are

\begin{align}
    \begin{split}
    \bra{n_1,\bm{k}_1;n_2,\bm{k}_2}\hat{V}\ket{n_3,\bm{k}_3;n_4,\bm{k}_4} &= \frac{1}{A}\sum_{\bm{q}\neq 0}v(\bm{q})\bra{n_1,\bm{k}_1}e^{i\bm{q}\cdot\bm{r}}\ket{n_3,\bm{k}_3}\bra{n_2,\bm{k}_2}e^{-i\bm{q}\cdot\bm{r}}\ket{n_4,\bm{k}_4}.
    \end{split}
\end{align}
Two-body interaction matrix elements in a generic magnetic Bloch band with index $i$ are then

\begin{align}
    \begin{split}
        \bra{i,\bm{k}_1;i,\bm{k}_2}\hat{V}\ket{i,\bm{k}_3;i,\bm{k}_4} &= \sum_{n_1,n_2,n_3,n_4}z^{*}_{in_1\bm{k}_1}z^{*}_{in_2\bm{k}_2}z_{in_3\bm{k}_3}z_{in_4\bm{k}_4}\bra{n_1,\bm{k}_1;n_2,\bm{k}_2}\hat{V}\ket{n_3,\bm{k}_3;n_4,\bm{k}_4}.
    \end{split}
\end{align}

\section{Hartree-Fock self energy}\label{sec:HF}

In this work, we consider many-body states in which the second miniband is partially occupied and the first miniband is full. Electrons in the second miniband acquire a Hartree-Fock self-energy through their interaction with electrons in the full first band. To show this, we consider the Hamiltonian of a generic multi-band system of electrons with a two-body interaction:
\begin{align}
    \begin{split}
        H &= \sum_{i,\bm{k}}\varepsilon_i(\bm{k})c^{\dag}_{i,\bm{k}}c_{i,\bm{k}} + \frac{1}{2}\sum_{ijlm}\sum_{\bm{k}'\bm{p}'\bm{k}\bm{p}}V_{i\bm{k}',j\bm{p}';l\bm{k},m\bm{p}}c^{\dag}_{i,\bm{k}'}c^{\dag}_{j,\bm{p}'}c_{m,\bm{p}}c_{l,\bm{k}}.
    \end{split}
\end{align}
Here $i$ is a band index and $\bm{k}$ is a crystal momentum label. $V_{i\bm{k}',j\bm{p}';l\bm{k},m\bm{p}} = \bra{i\bm{k}',j\bm{p}'}\hat{V}\ket{l\bm{k},m\bm{p}}$ is a two-body interaction matrix element. We assume a translation-invariant two body potential so that $V_{i\bm{k}',j\bm{p}';l\bm{k},m\bm{p}} \propto \sum_{\bm{g}}\delta_{\bm{k}'+\bm{p}',\bm{k}+\bm{p}+\bm{g}}$ where $\bm{g}$ is a reciprocal lattice vector. Consider a ``Fermi sea" single-Slater-determinant made of a full set of bands $\ket{\text{FS}} = \left(\prod_{i\in\text{full}}\prod_{\bm{k}}c^{\dag}_{i,\bm{k}} \right)\ket{0}$ where $\ket{0}$ is the vacuum state annihilated by all $c_{i,\bm{k}}$. Using Wick's theorem (see appendix 3 of Ref. \cite{giuliani2008quantum}), we can rewrite $H$ \emph{exactly} as
\begin{align}\label{eq:wickstheorem}
    \begin{split}
         H &= \sum_{i,\bm{k}}\varepsilon_i(\bm{k}):c^{\dag}_{i,\bm{k}}c_{i,\bm{k}}:  + \sum_{ij,\bm{k}}\Sigma_{ij}(\bm{k}):c^{\dag}_{i,\bm{k}}c_{j,\bm{k}}:  \\
         &+ \frac{1}{2}\sum_{ijlm}\sum_{\bm{k}'\bm{p}'\bm{k}\bm{p}}V_{i\bm{k}',j\bm{p}';l\bm{k},m\bm{p}}:c^{\dag}_{i,\bm{k}'}c^{\dag}_{j,\bm{p}'}c_{m,\bm{p}}c_{l,\bm{k}}: + \sum_{i \in \text{full},\bm{k}}\left(\varepsilon_i(\bm{k}) + \Sigma_{ii}(\bm{k}) \right)
    \end{split}
\end{align}
Here $``:\,:"$ denotes normal ordering with respect to $\ket{\text{FS}}$ (i.e. moving all ladder operators that annihilate $\ket{\text{FS}}$ to the right of those that do not, keeping track of minus signs coming from Fermi statistics). Also,
\begin{align}
    \begin{split}
        \Sigma_{ij}(\bm{k}) &= \Sigma^{H}_{ij}(\bm{k}) + \Sigma^{F}_{ij}(\bm{k}) \\
        &= \sum_{l\in \text{full}}\sum_{\bm{p}}\left( V_{i\bm{k},l\bm{p};j\bm{k},l\bm{p}} - V_{i\bm{k},l\bm{p};l\bm{p},j\bm{k}}\right)
    \end{split}
\end{align}
is the Hartree-Fock self-energy. Assuming $\ket{\text{FS}} = \left(\prod_{\bm{k}}c^{\dag}_{1,\bm{k}}\right)\ket{0}$ and retaining only terms for which the band index is $2$ (i.e. projecting to the second miniband), we find
\begin{align}
    \begin{split}
         \bar{H} &= \sum_{\bm{k}}\left(\varepsilon_2(\bm{k}) + \Sigma_{22}(\bm{k}) \right) c^{\dag}_{2,\bm{k}}c_{2,\bm{k}}\\
         &+ \frac{1}{2}\sum_{\bm{k}'\bm{p}'\bm{k}\bm{p}}V_{2\bm{k}',2\bm{p}';2\bm{k},2\bm{p}}c^{\dag}_{2,\bm{k}'}c^{\dag}_{2,\bm{p}'}c_{2,\bm{p}}c_{2,\bm{k}} + \sum_{\bm{k}}\left(\varepsilon_1(\bm{k}) + \Sigma_{11}(\bm{k}) \right).
    \end{split}
\end{align}
This is equivalent to Eq. 4 of the main text where $\tilde{\varepsilon}(\bm k) = \varepsilon_2(\bm{k}) + \Sigma_{22}(\bm{k})$ except for the final constant term $\sum_{\bm{k}}\left(\varepsilon_1(\bm{k}) + \Sigma_{11}(\bm{k}) \right)$.

We emphasize that $\bra{\text{FS}}A\bar{H}B\ket{\text{FS}} = \bra{\text{FS}}AH B\ket{\text{FS}}$ where $A$ and $B$ are arbitrary polynomials in the $c_{2,\bm{k}}$'s and $c^{\dag}_{2,\bm{k}}$'s, so that $A\ket{\text{FS}}$ and $B\ket{\text{FS}}$ are arbitrary states in which the first band is full and the second band is partially occupied. This is because all terms in $H$ (Eq. \ref{eq:wickstheorem}) are normal ordered with respect to $\ket{\text{FS}}$. All such terms involving non-second-band ladder operators thus annihilate $A\ket{\text{FS}}$. The only terms in $H$ that contribute to the above matrix element are those involving only second-band ladder operators or numbers. These terms appear both in $H$ and $\bar{H}$ and make identical contributions to both matrix elements. It follows that diagonalizing $\bar{H}$ in a basis spanned by the set of states $\{A\ket{\text{FS}}\}$ is equivalent to diagonalizing $H$ in that basis. In other words, \textbf{calculating the ground state of $\bar{H}$ in the basis $\{A\ket{\text{FS}}\}$ is equivalent to variationally minimizing $H$ in this basis}. Further, $\bra{\text{FS}}A\bar{H}B\ket{\text{FS}} = \bra{0}A\bar{H}B\ket{0}$, so we are free to think of our basis states as either $\{A\ket{0}\}$ or $\{A\ket{\text{FS}}\}$ for the purpose of calculation. The latter choice is more physically meaningful.

\section{Adiabatic model for twisted MoTe$_2$}
The adiabatic model for $t$MoTe$_2$ was originally derived in Ref. \cite{morales2023magic}. 
While Ref. \cite{morales2023magic} works in the electron picture, we work in the hole picture, as in Ref. \cite{reddy2023fractional}. 
The $K$-valley projected continuum model for AA-stacked TMD homobilayers is given in layer space by \cite{wu2019topological}
\begin{equation}
    H = \begin{pmatrix}
        \frac{1}{2m^*} (\bm p - \hbar\bm k_b)^2+ \Delta_b(\bm r) & \Delta_T(\bm r)\\
        \Delta_T(\bm r)^\dagger & \frac{1}{2m^*} (\bm p- \bm \hbar \bm k_t)^2 + \Delta_t(\bm r)
    \end{pmatrix}.
\end{equation}
The moir\'e potential and interlayer tunneling terms are given by

\begin{subequations}
\begin{align}
    \Delta_{b/t} &= -2 V\sum_{i=1,3,5} \cos(\bm g_j\cdot \bm r \pm \phi)\\
    \Delta_T(\bm r) &= w(1+e^{-i\bm g_2\cdot \bm r} + e^{-i\bm{g}_3\cdot \bm r}),
\end{align}
\end{subequations}
where $V,w$ and $\phi$ are parameters typically determined by fitting to DFT calculations. We adopt the parameters from Ref. \cite{reddy2023fractional} for $t$MoTe$_2$: $(V,w,\phi,m^*,a_0) = (11.2 \text{meV}, -13.3 \text{meV}, -91^\circ,0.62 m_e, 3.52 \mathring{A})$ and the moiré lattice constant is $a = a_0(2\sin(\theta/2))^{-1}$. The momentum shifts in the top and bottom layers are given by $\bm{k}_{b/t} = k_\theta(-1/2, \pm 1/2\sqrt{3})$ and the first shell of moir\'e reciprocal lattice vectors are $\bm{g}_j = k_\theta(\cos(\pi (j-1)/3),\sin(\pi (j-1)/3))$ for $j=1,\ldots, 6$, with $k_\theta = 4\pi/\sqrt{3}a$. The Hamiltonian for valley $K'$ is the time reversal conjugate of $H$. To map this model onto a skyrmion Chern band model as discussed in the main text, it is necessary to remove the layer-dependent momentum shifts in the kinetic terms. This is accomplished by a gauge transformation $H \rightarrow UH U^{\dag} = H'$ where
\begin{align}
    \begin{split}
        U = \begin{pmatrix}
            e^{-i\bm k_b \cdot \bm{r}} & 0 \\
            0 & e^{-i\bm k_t\cdot \bm{r}}
        \end{pmatrix},
    \end{split}
\end{align}
after which the tunneling becomes $\tilde \Delta_T(\bm{r}) = \Delta_T(\bm{r})e^{i(\bm{k}_b-\bm{k}_t)\cdot \bm r}$. The new Hamiltonian can be written
\begin{equation}
        H' = \frac{\bm{p}^2}{2m^*} + \bm{J}(\bm r) \cdot \bm {\sigma} + V(\bm r)
\end{equation}
where $\bm{J}(\bm{r}) = (\text{Re}(\tilde \Delta_T),\text{Im}(\tilde \Delta_T),\frac{\Delta_b-\Delta_t}{2})$ and $V(\bm r) = \frac{\Delta_b+\Delta_t}{2}$. $H'$ is now in skyrmion Chern band model form, except for an additional scalar potential term $V(\bm{r})$. The effective adiabatic Hamiltonian results from performing the same manipulations leading from Eq. (1) to Eq. (2) in the main text. Explicitly,
\begin{align}
    \begin{split}
  H_{\text{ad,TMD}} = \frac{(\bm p- e\bm A(\bm r))^2}{2m^*} +\frac{\hbar^2}{8m^*} (\partial_i \hat{\bm S})^2 - J(\bm {r}) + V(\bm{r})
  \end{split}
\end{align}
where $J(\bm{r})= |\bm J(\bm r)|$, $\hat{\bm{S}} = \bm{J}(\bm{r})/ J(\bm r)$, and $\nabla \times \bm A = -\frac{\hbar}{2e}\hat{ \bm S} \cdot (\partial_x \hat{\bm S} \times \partial_y\hat{ \bm S})$.

\begin{figure*}
    \centering
\includegraphics[width=0.5\textwidth]{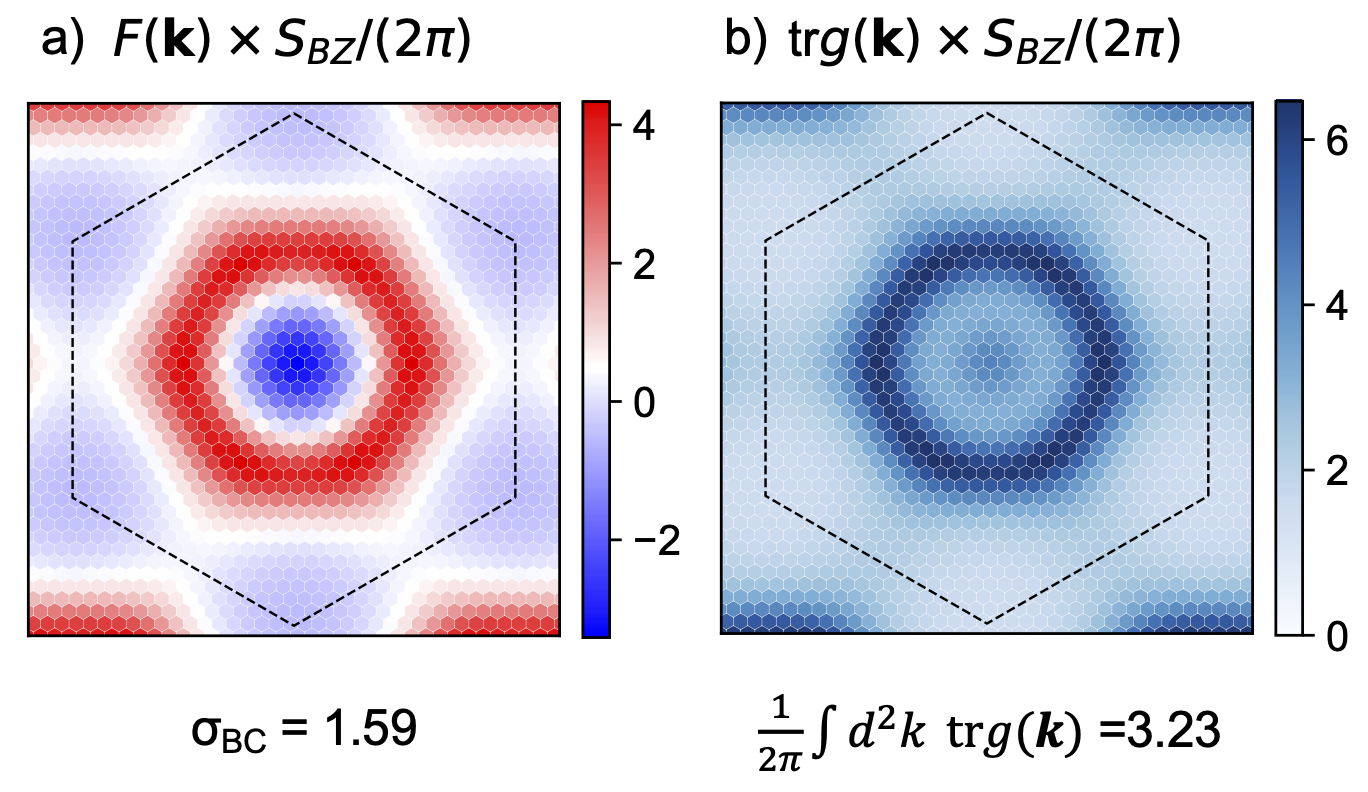}
    \caption{(a) Berry curvature $F(\bm{k})$ and (b) quantum metric trace $\text{tr} g(\bm k)$ of the second $t$MoTe$_2$ adiabatic model band at $\theta=2.5^{\circ}$. $\sigma_{BC}$ is the standard deviation of the Berry curvature, in units such that its average is unity. $S_{BZ}$ is the $k$-space area of the Brillouin zone.}
\label{fig:QGeomTMD}
\end{figure*}

\begin{figure*}
    \centering
\includegraphics[width=\textwidth]{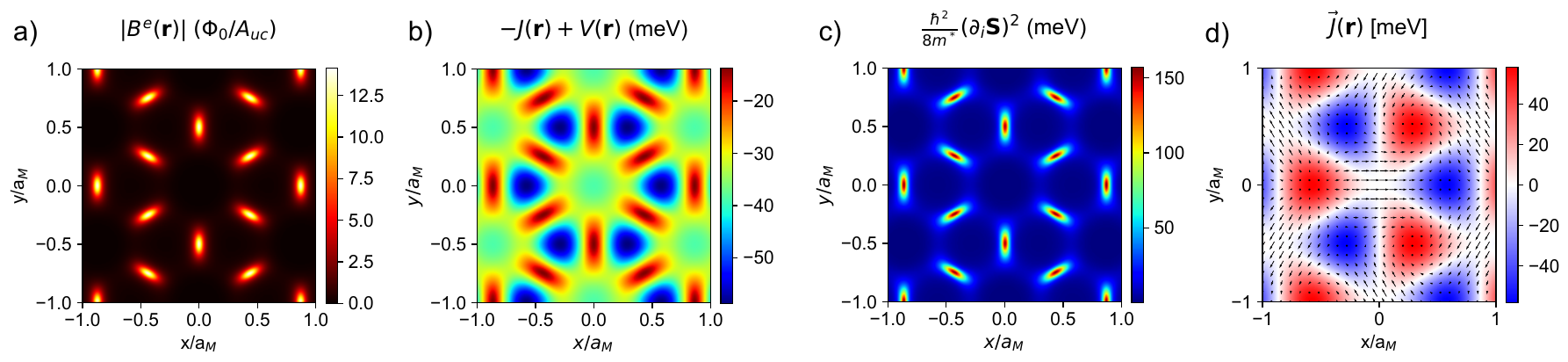}
    \caption{Emergent magnetic field (a) and the effective scalar potentials (b,c) entering $H_{\text{ad,TMD}}$. Panel (c) is evaluated at $\theta=2.5^{\circ}$ while (a) and (b) are independent of $\theta$. (d) Layer zeeman field in $H'$. Color represents $J_z$ and arrows represent $J_{x/y}$.}
\label{fig:adiabatic}
\end{figure*}

In Fig. \ref{fig:QGeomTMD}, we show the Berry curvature and the trace of the quantum metric of the second $t$MoTe$_2$ adiabatic model band at $\theta=2.5^{\circ}$. In Fig. \ref{fig:adiabatic}, we plot the various terms entering $H_{\text{ad,TMD}}$.

\bibliographystyle{apsrev4-1}